\documentclass[usenatbib,usegraphicx,useAMS]{mn2e}
\usepackage{aas_macros}
\usepackage{amssymb}
\usepackage{pdflscape}
\newcommand{\kms}{km s$^{-1}$}
\newcommand{\ergs}{erg s$^{-1}$}
\newcommand{\mics}{$\mu$m}
\newcommand{\msun}{M$_{\odot}$}

\newcommand{\lfuv}{L$_{\rm FUV}$}
\newcommand{\lir}{L$_{\rm IR}$}
\newcommand{\lirs}{$\bar{\textrm{L}}_{IR}$}
\newcommand{\sfrs}{$\overline{\textrm{SFR}}$}
\newcommand{\lirl}{L$_{\rm IR}^{l}$}
\newcommand{\dlir}{$\Delta_{\rm SFR}$}
\newcommand{\smass}{M$_{\star}$}
\newcommand{\othree}{[O III]$\lambda5007$}
\newcommand{\lothree}{L$_{\rm [O~III]}$}
\newcommand{\ntwo}{[N II]$\lambda6584$}
\newcommand{\oiii}{[O III]}

\newcommand{\hb}{H$\beta$}
\newcommand{\ha}{H$\alpha$}
\newcommand{\dfour}{$D_{n}4000$}

\title[SDSS galaxies in Herschel/Stripe 82]{Local SDSS galaxies in the Herschel Stripe82 survey: A critical assessment of 
optically-derived star-formation rates}

\author[Rosario et al.]
{D.J.~Rosario$^{1,2}$, J.T.~Mendel$^{1}$, S.L.~Ellison$^{3}$, D.~Lutz$^{1}$, J.R. Trump$^{4,5}$ \\
$^1$Max Planck Institute for Extraterrestrial Physics, Giessenbachstrasse 1, 85748 Garching, Germany\\
$^2$Department of Physics, Durham University, South Road, Durham DH1 3LE, UK\\
$^3$Department of Physics and Astronomy, University of Victoria, BC V8P 5C2, Canada\\
$^4$Department of Astronomy and Astrophysics, 525 Davey Lab, The Pennsylvania State University, University Park, PA 16802, USA\\
$^5$Hubble Fellow}

\begin{document}

\maketitle

\begin{abstract}
   
We study a set of 3319 galaxies in the redshift interval $0.04 < z < 0.15$ with 
far-infrared (FIR) coverage from the Herschel Stripe 82 survey (HerS), and emission-line measurements, 
redshifts, stellar masses and star-formation rates (SFRs) from the SDSS (DR7) MPA/JHU database. 
About 40\% of the sample are detected in the Herschel/SPIRE 250 \mics\ band. 
Total infrared (TIR) luminosities derived from HerS and ALLWISE photometry
allow us to compare infrared and optical estimates of SFR with unprecedented statistics for diverse classes of galaxies. 
We find excellent agreement between TIR-derived and emission line-based SFRs for H II galaxies. Other classes, 
such as active galaxies and evolved galaxies, exhibit systematic discrepancies between optical and TIR SFRs. 
We demonstrate that these offsets are attributable primarily to survey biases and the large intrinsic uncertainties of 
the \dfour-- and colour--based optical calibrations used to estimate the SDSS SFRs of these galaxies.
Using a classification scheme which expands upon popular emission-line methods, we
demonstrate that emission-line galaxies with uncertain classifications include a population of 
massive, dusty, metal-rich star-forming systems that are frequently neglected in existing studies. 
We also study the capabilities of infrared selection of star-forming galaxies. FIR selection reveals a substantial
population of galaxies dominated by cold dust which are missed by the long-wavelength WISE bands. 
Our results demonstrate that Herschel large-area surveys 
offer the means to construct large, relatively complete samples of local star-forming galaxies with 
accurate estimates of SFR that can be used to study the interplay between nuclear activity and star-formation.

\end{abstract}

\section{Introduction}

The Herschel Space Observatory \citep{pilbratt10}, now in its archival phase, has redefined our knowledge of dusty star-formation in our
Galaxy and out to the distant cosmos. As part of its legacy, Herschel targeted large areas of the sky for extragalactic
science. This capability was thanks in part to the sensitivity and large field-of-view of the SPIRE imaging instrument \citep{griffin10}, which 
quickly approached the confusion limit in its range of operating wavelengths ($250 - 500$ \mics) even with sky-scan rates as high
as 60" sec$^{-1}$. 

While much of the effort from Herschel extragalactic surveys have concentrated on galaxy populations in the distant Universe \citep{lutz14},
wide-area mapping programs are also suited to studies of more nearby galaxies. For example, the largest Herschel survey of the
extragalactic sky is the Herschel-ATLAS \citep{eales10}. With its total area of 550 deg$^{2}$, it spans a co-moving volume of 0.35 Gpc$^3$
out to $z=0.2$, approximately that of the largest dark matter simulations currently available, such as the Millennium simulation \citep{springel05}.  
As we will demonstrate, confusion-limited photometry of SPIRE images can detect the far-infrared (FIR) dust emission in the
majority of local massive star-forming galaxies from such surveys. In addition, the SPIRE PSF (FWHM $> 18$") is 
broad enough to allow the treatment of all but the very nearest galaxies as point sources, easing the process of photometric characterisation. 
But it is also narrow enough to prevent significant blending of galaxies even in compact groups. 
Therefore Herschel wide-area photometric catalogs combined with local galaxy
surveys, such as the SDSS Main Galaxy sample, offer an ideal dataset to explore the star-formation and dust properties of 
normal galaxies over cosmologically interesting volumes. 

The FIR also has unique value as a reliable tracer of the star-formation rate (SFR) in active galaxies, in which the emission
from active galactic nuclei (AGN) can otherwise significantly contaminate or overpower optical, UV or emission line tracers.
The long-wavelength FIR ($>100$ \mics) has been demonstrated to arise primarily from dust that is unaffected by the intense
radiation from the nucleus, even in galaxies that host luminous AGN \citep{netzer07, mullaney11a, rosario12}. Studies of star-formation
in active galaxies from the SDSS generally rely on the \dfour\ calibration of the SFR or others that involve spectral indices (Section 2.2). 
These can be critically appraised and, if needed, replaced by direct calorimetric tracers such as the FIR luminosity.

Despite these motivations, only few studies in the existing literature have looked at the Herschel view of the local Universe beyond the very nearest, 
resolved systems which have been targeted individually in large programs such as HRS \citep{boselli10} and KINGFISH \citep{kennicutt11}. 
\citet{lam13} studied the relationships between total infrared (TIR) and \ha\ luminosities for $\approx 290$ emission line galaxies from the SDSS 
in the 14 deg$^2$ science demonstration release of the Herschel-ATLAS. These two tracers
were well correlated, though the tightness of the correlation grew increasingly worse towards longer FIR wavelengths. 
At high emission line luminosities, active galaxies deviated from the typical correlations due to the component of AGN-ionised gas.

With another modest sample of SDSS galaxies in smaller area deep extragalactic fields ($\approx 2.2$ deg$^2$), \citet{dominguez-sanchez14} 
used multi-wavelength tracers, including Herschel/PACS 100 and 160 \mics\ photometry, to systematically compare
the performance of various star-formation indicators, finding broad consistency between them. They also examined the location
of Herschel-detected galaxies on the plane of star-formation rate (SFR) and stellar mass (\smass), confirming that these galaxies
lie on the so-called Main Sequence of star-formation, a ridgeline feature 
in this plane which defines the location of most star-forming galaxies \cite[e.g.][]{noeske07,elbaz07,salim07,speagle14}.

While breaking new ground, these studies are limited by their small sample sizes, which especially limits conclusive results
for rare populations, such as active galaxies. In addition, a full contextual study of FIR flux-limited samples also requires a treatment
of the galaxies that are not detected in the FIR.

In this work, we study the properties of $\approx 3300$ galaxies from the SDSS that lie within 
the Herschel Stripe 82 survey, including a full characterisation of the 
subset of about 1300 galaxies that are detected in the FIR. We examine their redshift distributions, classifications, TIR luminosities and 
detection rates, and compare optical- and infrared-based SFR calibrations, including the important limitations set by the large systematic uncertainties
on SFRs from spectral index and SED-fitting techniques. This study is the first part of a two paper series on the star-forming properties
of narrow-line selected AGN in the SDSS and serves to introduce our dataset and key measurements that will be used in the second
paper. 
  
We adopt a standard $\Lambda$-CDM Concordance cosmology, with $H_{0} = 70$ \kms~Mpc$^{-1}$ and $\Omega_{\Lambda}=0.7$. 
Stellar masses reported in this study are estimated with a \citet{chabrier03} IMF. Luminosities are quoted in \ergs\ and SFRs in
\msun/yr, unless otherwise stated.

\section{Datasets and Methods}

In this section, we describe the datasets and methods used for the analyses in the rest of this work.
We draw on the extensive set of measurements (redshifts, stellar masses, line fluxes, etc.~) available from the SDSS/MPA-JHU
database to define and classify our sample of galaxies for study, and to understand their range of observed properties.
Where possible, we employ Herschel and WISE photometry to constrain the TIR using SED fits,
and GALEX FUV photometry to correct for stellar emission not absorbed by dust. These data are brought
to bear on the cross-assessment of infrared (IR), ultraviolet (UV) and optical SFRs in later sections.

\subsection{The Herschel Stripe 82 survey}

The Herschel Stripe82 survey \citep[HerS;][]{viero14} consists of 250, 350 and 500 \mics\ Herschel/SPIRE maps over a total area of 79 deg$^2$
in the RA range of 13$^{\circ}$ to 37$^{\circ}$ and a Declination range of -2$^{\circ}$ to 2$^{\circ}$.
The field was covered using two sets of incompletely overlapping, nearly orthogonal scans, leading to some variation in depth. 
The large SPIRE PSF (FWHM of 18.2'', 25.2'', 36.3'' at 250, 350 and 500 \mics\ respectively) ensures that almost 
all galaxies are point sources in HerS. The maps are deep enough to approach the confusion limit at 250 \mics, and 
confusion noise is a substantial factor in the source detection limit of the survey.

We make use of the public survey catalog available from the HerS website \footnote{http://www.astro.caltech.edu/hers/HerS\_Home.html}.
A detailed treatment of the source detection and characterisation is available in \citep{viero14}, and we summarise main aspects here.
The catalog is 250 \mics\ selected to a $3\sigma$ depth of 28-31 mJy, depending on the local coverage of the map. 
Source detection was performed using the STARFINDER package
\citep{diolaiti00} on a high-pass filtered version of the 250 \mics\ map, designed to remove extended Galactic emission while 
not strongly influencing the PSF of point sources. Nearby extended sources were also removed from this catalog. The DESPHOT method
of \citet{roseboom10} optimised for SPIRE was used for photometry of the detected sources in all bands. This technique
deals with deblending of SPIRE sources effectively and uses 250 \mics\ sources as priors for photometry from the lower resolution
350 and 500 \mics\ bands. As a figure-of-merit, the approximate 250 \mics\ depth of $30$ mJy is sufficient to detect a galaxy at $z=0.08$ 
with a SFR of $1$ \msun/yr, making it a valuable tool to study obscured and dusty star-formation in the local Universe.

\subsubsection{Infrared SED templates}

Galaxies vary in the relative amounts of warm and cold dust \citep[e.g.,][]{dale01,dale02,ciesla14}, which affects the
MIR-to-FIR luminosity in their infrared (IR) spectral energy distributions (SEDs). \citet{dale02} present a family of IR
SED templates that encapsulate much of this variation among galaxies in the local Universe. The library
of templates is parameterised by $\alpha$, the power-law index of the radiation intensity ($U$) incident on the dust grains.
From \citet{dale01}, $\alpha$ is given by:

\begin{equation}
\mathrm{d} M(U)  \; = \; U^{-\alpha} \; \mathrm{d}U
\end{equation}

\noindent where $M(U)$ is the mass of dust heated by radiation of intensity U. We work with a subset of 48 templates with
$0 < \alpha \leq 3$, since the rest of the templates are indistinguishable from the template with $\alpha = 3$ in the IR
bands used for our SED fits.

$\alpha$ is closely related to the temperature distribution of the grains. Approximating the \citet{dale02} SEDs as modified black bodies with a characteristic
dust temperature $T_{dust}$, \citet{magnelli14} have calculated a monotonic conversion between $\alpha$ and $T_{dust}$.
Readers interested in the dust temperatures of galaxies in this work can consult Figure \ref{alpha_dist} in which both $\alpha$
and $T_{dust}$ are presented.

\subsection{SDSS measurements}

By design, HerS overlaps with the Stripe 82 of the Sloan Digital Sky Survey (SDSS) % (Figure \ref{field_map}).
over about 70 deg$^2$.
The Stripe 82 was the target of multiple imaging passes in the five SDSS bands. The imaging is considerably deeper than over 
the regular survey, making the Stripe 82 particularly useful for detailed structural studies of galaxies in the Local Universe. 
The SDSS spectroscopic survey covers Stripe 82 with high areal sampling completeness 
($> 90$\%), and the SDSS spectrograph has provided moderate resolution spectra of the central parts of galaxies
through the 3'' diameter of its fibre apertures. 
We consider galaxies from the SDSS Main Galaxy sample 
\citep{strauss02}, with $r_{AB} < 17.77$ and confirmed redshifts in the range of $0.04 < z < 0.15$. 
The redshift interval covers the majority of SDSS spectroscopic targets
but is small enough to keep redshift-dependent selection and aperture effects manageable, 
and reduces the need to account for galaxy evolutionary trends which become non-negligible by $z\sim0.3$. 
A total of 3319 SDSS galaxies that satisfy the main selection criteria lie in the HerS footprint, and we refer to them collectively
as the working sample. For reference, the comoving volume spanned by our sample is $2\times10^{7}$ Mpc$^3$, considerably
larger than those of modern cosmological hydrodynamic simulations such as Illustris \citep{vogelsberger14} or EAGLE \citep{schaye15}.

We adopt emission line fluxes, aperture corrections, stellar masses and star-formation rates (SFRs) from the
MPA-JHU SDSS Data Release 7 \citep[DR7;][]{abazajian09} database\footnote{http://www.mpa-garching.mpg.de/SDSS/DR7/} of galaxy measurements. 
We summarise the methodology for estimating these quantities, but further details may be found on the database webpages and in related
publications \citep{kauffmann03a, brinchmann04, salim07}.

Line fluxes were measured from the SDSS spectra after subtraction of a model for the stellar continuum of the galaxy. 
For those galaxies with line spectra dominated by the emission from H II regions (i.e., H II galaxies), 
calibrations based on multi-line ionisation models were used to determine the dust extinction-corrected SFR within the fibre.
In other galaxies, such as those with weak line fluxes or those influenced by ionisation from AGN or other sources besides young stars, 
a different calibration was employed to derive fibre SFRs, based on a relation between the line-based 
specific star-formation rate (sSFR) and the optical \dfour\ index in H II galaxies. We will use the term `\dfour\ calibration'
to refer to this method later in the paper.

The SDSS photometry beyond the fibre aperture was calculated by subtracting 
simulated photometry extracted from their spectra. The SFRs of all galaxies beyond the fibre
were estimated from this photometry in a uniform fashion, using SED models calibrated against 
UV-optical tracers based on the study of \citet{salim07}.
The total SFRs of a galaxy is the sum of the individual fibre and outer SFRs. 
% Since different calibrations
%were used for the fibre SFRs of AGN and H II galaxies, there may be systematic differences between these estimates. However,
%the different methods were calibrated to agree in a statistical sense \citep{brinchmann04}. In addition, 
The total SFR is usually much higher than the fibre SFR -- the median ratio is $\approx 5$ with considerable scatter. 
The total SFRs are mostly influenced by the component outside the fibre, which is estimated using the same methodology for all classes of galaxy.

Stellar masses were estimated from the total SDSS five-band photometry using the same SED models used to estimate the SFRs, both
within and outside the fibre aperture. The MPA-JHU database specifically excludes galaxies with clear broad AGN lines; 
therefore, all AGN in this study are ``narrow-line" systems, i.e., their nuclear continuum emission is strongly obscured. This ensures
that the stellar mass estimates are not significantly contaminated by AGN light. Corrections for the emission line fluxes were applied
to the fibre photometry before estimating the fibre masses, so the high equivalent width of narrow emission lines from ionised
regions around powerful AGN do not bias their masses.

SDSS sources were linked to the HerS catalog using a simple closest distance crossmatch with a maximum tolerance of 5''. The
choice of this tolerance was based on the distribution of separations between HerS and SDSS sources singly matched
with a circular tolerance of 20'', larger than the 250 \mics\ PSF size. From the working sample, 1349 galaxies 
are matched to a 250 \mics--selected HerS counterpart, with no multiple matches or duplicate matches; a FIR detection rate of 40\%.
Of these, 756 galaxies are detected at 350 \mics, and 165 galaxies are also detected at 500 \mics.

\subsection{WISE 22 \mics\ photometry}

% ************************* Fig 1 ***********************************
\begin{figure}
\includegraphics[width=\columnwidth]{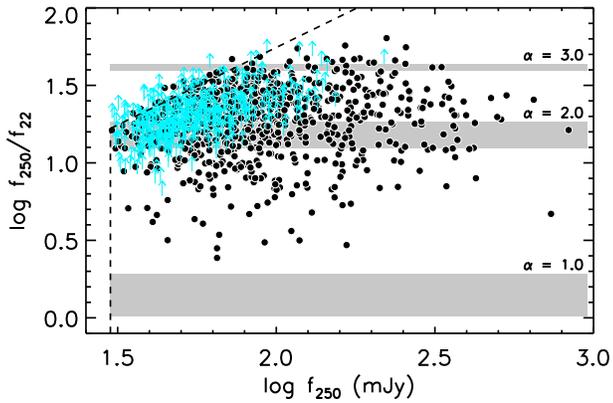}
\caption{ The ratio of the HerS 250 \mics\ and ALLWISE 22 \mics\ flux densities
plotted against the HerS 250 \mics\ flux density. Sources detected
in both bands are plotted as black points (SNR$>3$ at 250 \mics\ and SNR$>2$ at 22 \mics). 
The dashed lines establish the parts of the diagram censored by the median flux limits of the datasets. 
Grey bands show the ranges in the flux ratio spanned by three different galaxy IR SEDs from the \citet{dale02} 
library, where the thickness of the bands arise from k-corrections over
the redshift range $0.04<z<0.15$. The three SEDs are identified by their
$\alpha$ parameter (see text for details). The limits imply that some 
fainter HerS-detected galaxies with IR SEDs dominated by cold dust emission
(high $\alpha$) are not detected to the limit of the ALLWISE W4 photometry.
}
\label{mir_fir_color}
\end{figure}
% ************************* Fig 1 ***********************************

The Wide-field Infrared Survey Explorer \citep[WISE,][]{wright10} has mapped the entire sky in four MIR bands at
3.4 \mics, 4.6 \mics, 12 \mics\ and 22 \mics. A public catalog is available consisting of
photometry from the all-sky survey atlas of all point-like and extended sources detected with a S/N$>5$ 
in at least one of the four WISE bands. The catalog is not uniform, since the depth of the WISE imaging 
varies considerably across the sky. The astrometry accuracy of the catalog, tied to the 2MASS coordinate system, 
is better than 200 mas. 

We crossmatched the working sample with the ALLWISE all-sky survey catalog available from the IPAC/IRSA service, using a simple
cone search with a tolerance of 2''. About 96\% have a counterpart in the WISE survey, consistent
with the success rate of other studies \citep[e.g.,][]{rosario13d}.
We rely on the photometry performed by the ALLWISE survey pipeline, details of which may be found 
in the Data Release Supplement (http://wise2.ipac.caltech.edu/docs/release/allwise/expsup/).
The pipeline provides several different photometric measurements for sources in all four WISE bands. The primary photometry
is performed by PSF decomposition of sources, which assumes that pipeline targets are
composed of blends of point sources (the `profile-fitting' or PRO photometry). A maximum-likelihood model 
of the source plane is simultaneously fit in all four bands for contiguous batches of sources, while as many as two PSF 
components are allowed per source. In addition, various flavours of circular and elliptical aperture photometry are also provided by the survey pipeline.

In this work, we only employ photometry in the longest wavelength W4 band (22 \mics), covering a continuum-dominated part of the thermal
IR beyond the bands emitted by polycyclic aromatic hydrocarbons \citep{rosario13d}. 
The FWHM of the WISE PSF in the W4 band is $\approx 11\farcs8$. We use the ALLWISE
profile-fit $\chi^2$ at 22 \mics\ as a way to identify if the image of a source in that band is likely to be extended
\citep[see the Data Release Supplement and the discussion in][]{rosario13d}. All the galaxies in our
working sample are consistent with being point sources in the W4 band, allowing us to adopt their PRO
photometry. 830 galaxies from the HerS-detected working sample ($\approx 60$\%) are detected in W4 with SNR$\geq 2$. This SNR
threshold is lower than what one may traditionally adopt for detection, but since we only use WISE photometry for sources
with a firm 250 \mics\ prior detection, we use the lower significance measurements from ALLWISE for our SED fits (Section 2.4). 

The WISE pipeline also provides an estimate of the W4 flux limit for every source based on the
results of the PSF decomposition. 
These limits have typical value of 2.5 mJy (3$\sigma$), but there is a considerable tail of shallower limits
as a consequence of confusion in the WISE atlas images.

While the flux density limit of the 22 \mics\ catalog is about an order
of magnitude deeper than HerS at 250 \mics, the detection rate of our working sample in ALLWISE W4 is
considerably lower than in HerS. The reason for this can be appreciated through Figure \ref{mir_fir_color}.
At the W4 flux limit,
a population of galaxies with $f_{250} < 0.1$ Jy and high $\alpha$ (cold dust-dominated SEDs) are missed in ALLWISE
but detectable in HerS. This effect also presents an important corollary relevant for MIR studies of local galaxies \citep[e.g.,][]{lee13}:
galaxy samples selected purely from the thermal MIR are biased towards warm dust SEDs, while 
selections based on FIR photometry over the Rayleigh-Jeans (R-J) tail of the cold dust emission, 
span the full range in dust SED shapes. 

\subsection{GALEX FUV photometry}

The Herschel and WISE datasets sample the thermal IR SEDs and measure the dust-reprocessed UV and optical luminosity
of the SDSS galaxies in HerS. However, a fraction of the UV output from star-formation does escape unextinguished, which may be recovered
from direct UV photometry. In this work, we use GALEX far-UV (FUV) data from the Medium Imaging Survey (MIS) which covers a substantial part
of the Stripe82. The SDSS/DR7-GALEX/GR7 cross-matched subcatalog, available from the GALEX CasJobs SQL interface 
(http://galex.stsci.edu/casjobs/), was queried to recover all GALEX sources over the HerS footprint with an SDSS counterpart.
Associations were then made to the galaxies in our sample using the unique SDSS ObjIDs. To minimise spurious matches,
we applied a somewhat tighter cross-match tolerance of 3" over that of the public SDSS-GALEX catalog (which matches counterparts out to 5").  
 
In our analysis, we employ photometry from the GALEX FUV channel which covers the 1350--1750 \AA\ band, with a peak sensitivity 
at $\approx 1500$ \AA. Light at these wavelengths is particularly representatives of the emission from hot stars and this band has been 
successfully used to study unobscured star-formation in local galaxies \citep[e.g.][]{salim07, hao11}.

The GALEX coverage over the Stripe82 area is very non-uniform. When comparing detection statistics of SDSS galaxies in 
GALEX and HerS (Section 4), we restrict ourselves to galaxies that lie in regions with GALEX exposures $> 1$ ksec. We
determine these regions using a coverage map generated using the full MIS source catalog in the HerS footprint (also obtained
from GALEX CasJobs), binned in $1' \times 1'$ boxes over the field. This approach ensures that varying FUV depths do not lead to 
inconsistent conclusions about the relative amount of obscured and unobscured star-formation.
 
Of the 3066 galaxies of the working sample with adequate GALEX coverage, 1832 (60\%) are detected in the FUV band. 
K-corrections in the band are expected to be small ($\lesssim 30$\% to $z=0.15$ 
for the typical UV slope of galaxies).  Therefore, we calculate FUV luminosities (\lfuv) directly using the measured FUV fluxes.
 
% ************************* Fig 2 ***********************************
\begin{figure}
\includegraphics[width=\columnwidth]{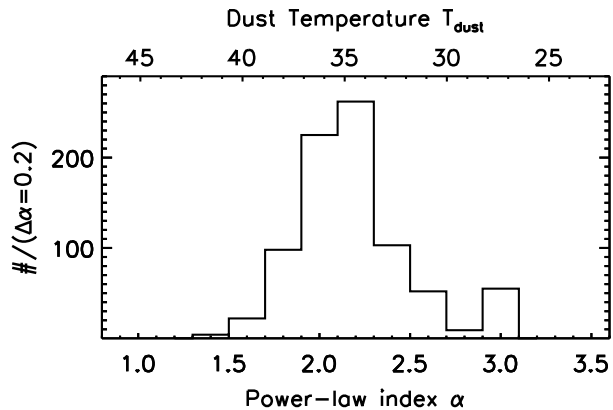}
\caption{ Distribution of the \citet{dale02} shape parameter $\alpha$ 
of the best-fitting SED template to galaxies from the working sample. High values of
$\alpha$ correspond to galaxies with colder typical dust temperatures. The median
value of $\alpha=2.2$ is typical for normal star-forming galaxies \citep{dale01}.
}
\label{alpha_dist}
\end{figure}
% ************************* Fig 2 ***********************************

In Table \ref{photometry_table}, we tabulate the Herschel/SPIRE, GALEX/MIS and WISE W4 photometry for all 3319
galaxies in the working sample. A minimum SNR of 3 is applied to all Herschel photometry, and a minimum SNR of 2
is applied to the WISE photometry. We do not apply a SNR threshold to the GALEX photometry, depending instead on the
criteria used in the MIS catalog. However, we do apply a minimum GALEX coverage as discussed above.

\begin{table*}
\label{photometry_table}
\centering
\caption{Photometry of galaxies from the SDSS Main Galaxy sample in the Herschel Stripe 82 survey at $0.04<z<0.15$ (Full version available with the online Journal)}
\begin{tabular}{ccccccccccc}
\hline
SDSS OBJID & $f_{\rm FUV}$ & $e_{\rm FUV}$ & $f_{\rm 22}$ & $e_{\rm 22}$ & $f_{\rm 250}$ & $e_{\rm 250}$ & $f_{\rm 350}$ & $e_{\rm 350}$ & $f_{\rm 500}$ & $e_{\rm 500}$  \\
 
 &(log Jy) & (log Jy) & (log Jy) & (log Jy) & (log Jy) & (log Jy) & (log Jy) & (log Jy) & (log Jy) & (log Jy)  \\
\hline 
588015509277573268 &  --   &  --   &  --   &  --   &  --   &  --   &  --   &  --   &  --   &  --  \\
588015507667419322 &  --   &  --   &  --   &  --   &  --   &  --   &  --   &  --   &  --   &  --  \\
588015507667419317 &  --   &  --   &  --   &  --   &  --   &  --   &  --   &  --   &  --   &  --  \\
587731511538876510 & -4.27 & -5.67 & -2.66 & -3.05 & -1.23 & -1.97 &  --   &  --   &  --   &  --  \\
587731513148047431 &  --   &  --   &  --   &  --   &  --   &  --   &  --   &  --   &  --   &  --  \\
587731512069587131 & -4.83 & -5.92 & -2.15 & -2.76 & -0.71 & -1.98 & -1.06 & -1.99 & -1.48 & -1.97\\
587731512610324707 &  --   &  --   & -1.57 & -2.89 & -0.29 & -1.99 & -0.63 & -1.98 & -1.18 & -1.95\\
587731513142542498 &  --   &  --   & -1.99 & -2.76 & -0.81 & -1.98 & -1.18 & -1.99 &  --   &  --  \\
587731512608292991 & -4.62 & -5.63 & -2.25 & -3.03 & -0.68 & -1.97 & -1.04 & -1.98 & -1.46 & -1.94\\
587731512069455963 &  --   &  --   & -2.33 & -3.05 & -0.83 & -1.96 & -1.10 & -1.98 & -1.36 & -1.95\\
587731512069587156 &  --   &  --   & -2.37 & -3.01 & -1.05 & -2.00 & -1.39 & -2.00 &  --   &  --  \\
587731513144377525 &  --   &  --   &  --   &  --   &  --   &  --   &  --   &  --   &  --   &  --  \\
587731514222510237 & -3.69 & -5.62 & -1.88 & -2.95 & -0.63 & -1.97 & -0.99 & -1.99 & -1.44 & -1.94\\
587731514222248063 &  --   &  --   &  --   &  --   &  --   &  --   &  --   &  --   &  --   &  --  \\
587731514222182633 & -3.60 & -5.56 & -1.90 & -3.02 & -0.47 & -1.99 & -0.73 & -1.99 & -1.11 & -1.97\\
\hline \hline

\end{tabular}
\end{table*}

\subsection{Infrared luminosities}

In this work, the total infrared luminosity (TIR) of a galaxy is defined as its integrated IR luminosity between 8 and 1000 \mics.
There are different standards in the literature for the range of wavelengths over which the TIR luminosity is integrated. 
For example, the recent review of \citet{kennicutt12} define the TIR over 3--1100 \mics. We will
instead use the older definition from \citet{kennicutt98}, which is more widespread in the existing literature. 
The MIR between 3 and 10 \mics\ can be affected in complex ways by emission from stellar photospheres
and very hot dust from star-formation and AGN \citep[e.g.][]{lu03}, and is therefore harder to constrain empirically.
 
The difference in the IR luminosity over 3--1100 \mics\ and 8--1000 \mics\ for the library of \cite{dale02} templates used in this work
ranges between 4\% and 15\%, and is therefore only a minor part of the TIR. We adopt a fixed difference of 10\%\ (0.041 dex) 
to allow a conversion between these two standards when exploring the performance of multi-wavelength SFR calibrations
in Section 4. This value is the median difference shown by galaxies from our working sample
with constrained SED shapes based on the fits discussed below.

The TIR luminosities (\lir) of galaxies are estimated using fits to their thermal IR SEDs, as sampled by the SPIRE bands
and the WISE 22 \mics\ band. The SEDs are fit to the library of IR galaxy templates from \cite{dale02} using the 
the Levenburg-Markquardt least-squares algorithm as implemented in the IDL routine MPFIT.  

Fits to SPIRE photometry alone do not allow any satisfactory discrimination of the full SED shape, 
since these bands span the R-J tail of dust emission. For the subset of galaxies with detections in both
the SPIRE bands and WISE 22 \mics\ band, the larger baseline in wavelength gives more accurate constraints on
the dust SED. For these galaxies, we estimated the uncertainty on \lir\ as the range in luminosities shown by IR templates 
that differed from the best-fit template by a reduced $\bar{\chi^{2}} < 1$. This systematic term to the \lir\ uncertainty is generally 
larger than the pure statistical uncertainty from the IR photometry.

A histogram of best-fit $\alpha$ for HerS- and ALLWISE-detected subset of the working sample is shown is 
Figure \ref{alpha_dist}. It peaks sharply at $\alpha=2.2$ or $T_{dust}=35$K, typical of normal star-forming galaxies \citep{dale01}. 
%There are very few warm, IR-luminous galaxies with low values of $\alpha$, which are likely missed because the volume occupied by the survey 
%does not adequately sample the bright end of the IR luminosity function. 
A small fraction of sources have a best-fit $\alpha=3$, an extreme value. There is very little variation 
in the long-wavelength shape of cold dust-dominated SEDs in the library and photometric
scatter leads to some sources that are not well constrained in terms of their $\alpha$ values. However, \lir\
is only weakly dependent on $\alpha$ in this regime and it remains robust despite the uncertainty in $\alpha$. 

We have demonstrated in Figure \ref{mir_fir_color} that galaxies without WISE 22 \mics\ detections are likely to have cold SED
shapes with characteristic $\alpha > 2$. To estimate \lir\ in this subset, we fit the SPIRE photometry to a fixed template with
$\alpha = 2.2$. The systematic uncertainty is calculated as the span in \lir\ of the best-fit SEDs with $\alpha$ in the range $2-2.5$, a 
factor of 0.25 dex.

\begin{figure}
\includegraphics[width=\columnwidth]{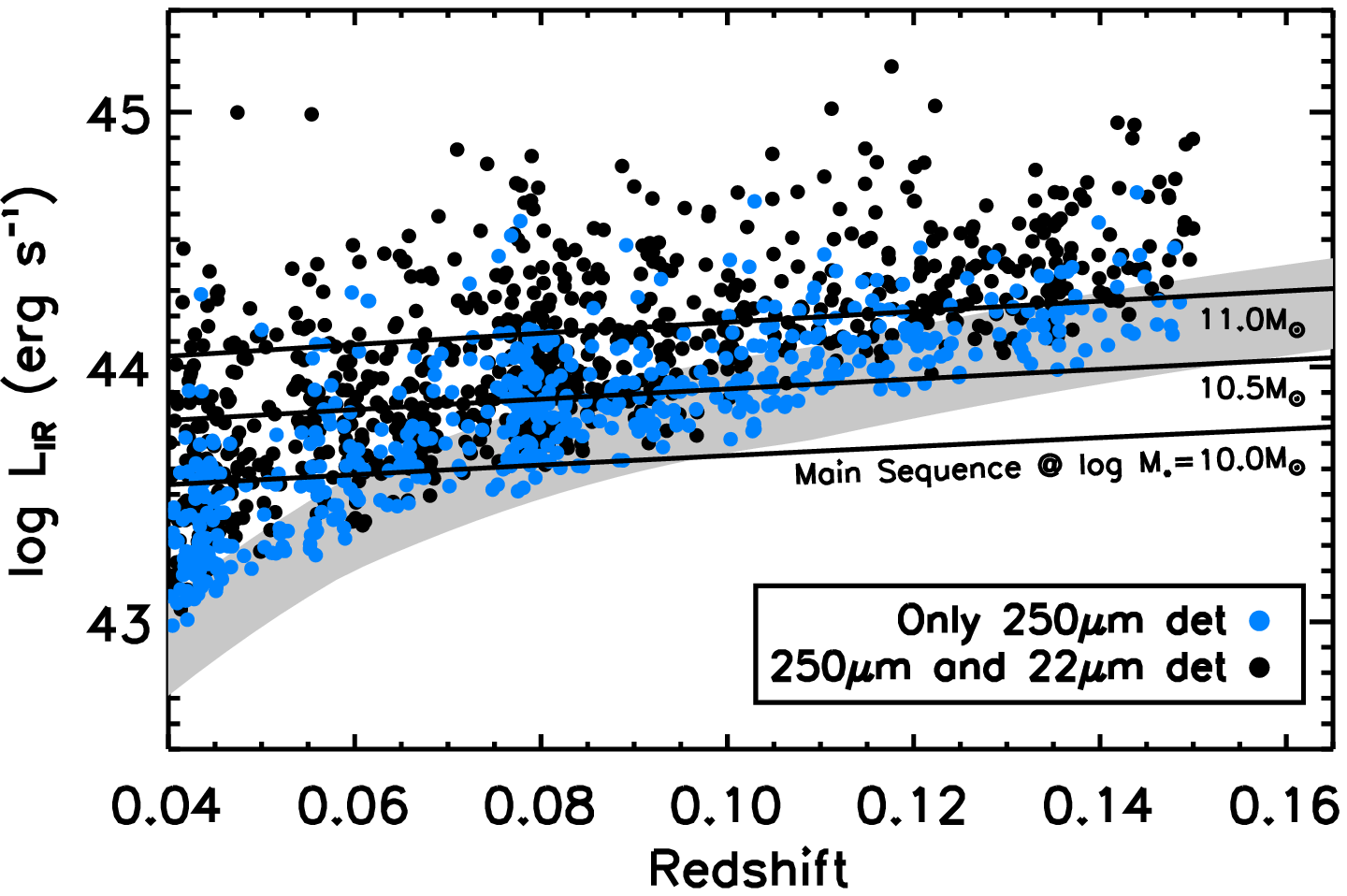}
\caption{8-1000 \mics\ luminosities (\lir) of HerS-detected galaxies from the working sample ($0.04<z<0.15$). Black points
show galaxies detected in the WISE $22$ \mics\ band, while blue points show galaxies selected only at $250$ \mics. The
grey band indicates the range of \lir\ of \citet{dale02} galaxy IR templates with $\alpha = [1.8, 2.5]$ with a 250 \mics\ flux at the 
characteristic HerS limit of 30 mJy. Sold black lines mark the ridge-line of the star-forming galaxy Main Sequence from \citet{speagle14}
at three fiducial stellar masses.}
\label{lir_vs_z}
\end{figure}

We plot our \lir\ estimates against redshift in Figure \ref{lir_vs_z}, distinguishing between galaxies detected at WISE 22 \mics\ (black points)
and those with only FIR detections (blue points). The \lir\ are consistent with the range of limiting IR luminosities (grey band) 
calculated with the 250 \mics\ flux limit of $30$ mJy and SEDs spanning the width of the $\alpha$ distribution in Figure \ref{alpha_dist}.
We also plot in the Figure a set of lines which mark the ridge-line value of the Main Sequence at three characteristic stellar masses
following the parameterisation of \citet{speagle14}. To $z\approx 0.8$, HerS can detect normal star-forming galaxies with masses
lower than $10^{10}$ \msun. Towards $z=0.15$, HerS-detected galaxies are increasingly dominated by starbursts. 
For this reason that we restrict our working sample to redshifts below 0.15.

% ************************* Fig 2 ***********************************
% ************************* Fig 2 ***********************************

\subsection{Galaxy classification}

% ************************* Fig 3 ***********************************
\begin{figure*}
\includegraphics[width=\textwidth]{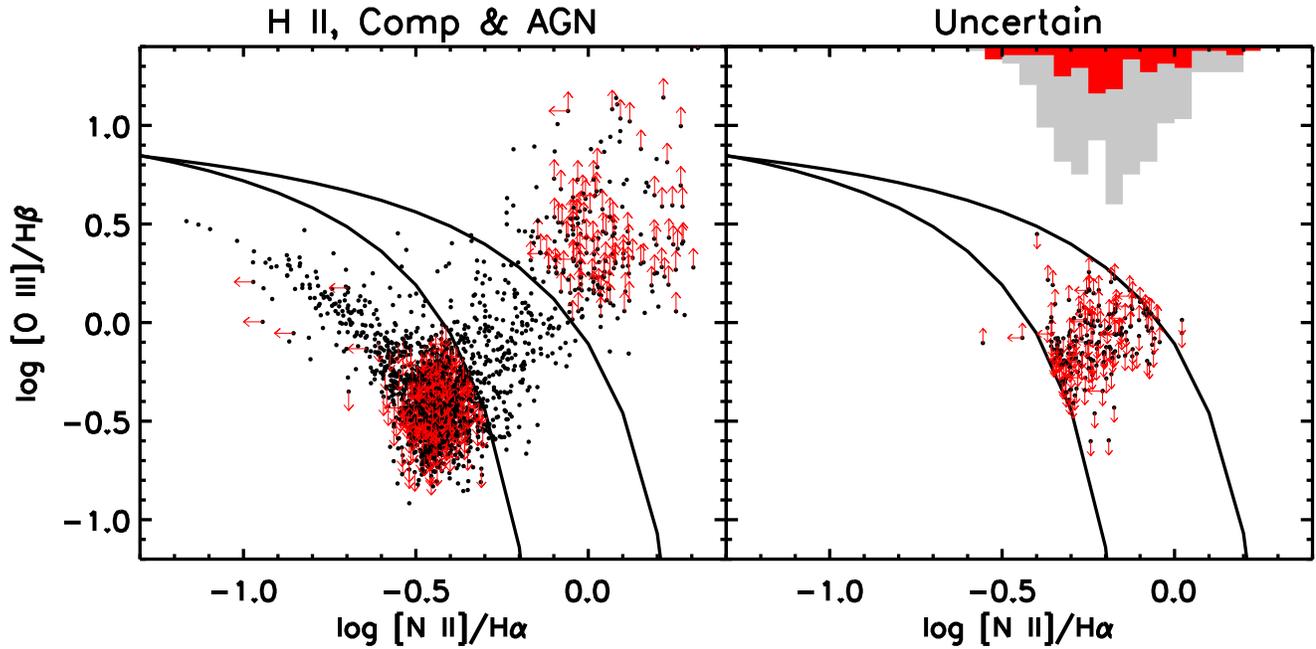}
\caption{The Baldwin-Phillips-Terlevich (BPT) diagram plotting the ratio of [N II]$\lambda 6584$/H$\alpha$ to the ratio 
of [O III]$\lambda 5007$/H$\beta$.The curved lines in both panels show the divisions between different classes of galaxies
following the formulae from \citet{kewley06}. In the left panel, we show galaxies from our working sample, 
classified either as H II, Composites or AGN. Red arrows mark galaxies with a limit in one of the two line ratios, with the direction
of the arrows indicating the nature of the limit. A number of galaxies are not detected in either the [O III] or H$\beta$ lines, but can still
be reliably classified based on the 3$\sigma$ limit on the [O III]/H$\beta$ ratio. The right panel shows only the galaxies classified
as Uncertain. Those for which a definite limit on the [O III]/H$\beta$ ratio can placed are plotted as small black points with the nature
of the limit shown with red arrows. The remainder of this population are undetected in both \othree\ and \hb, and cannot be plotted as points.
The distribution of their [N II]/\ha\ ratio is shown using the inverted histograms along the top X-axis (grey if [N II] is detected, limits on the
ratio in red if [N II] is undetected).
}
\label{bpt_byclass}
\end{figure*}
% ************************* Fig 3 ***********************************

\begin{table*} 
\begin{minipage}{\textwidth}
\centering
\caption{Classifications of galaxies}
\begin{tabular}{lcccccc}
\hline
Redshift bin & Dim & H II & Uncertain & Composite & AGN & Total \\
\hline \hline
\multicolumn{7}{c}{SDSS working sample} \\
\hline
$0.04<z<0.10$ & 697  & 955  & 302  & 173  & 153  & 2280 \\
$0.10<z<0.15$ & 434  & 273  & 210  & 70   & 52   & 1039 \\
\hline
\multicolumn{7}{c}{SDSS working sample - 250 \mics\ detected} \\
\hline
$0.04<z<0.10$ & 28  & 631  & 128  & 122  & 56   & 965  \\
$0.10<z<0.15$ & 15  & 213  & 95   & 44   & 17   & 384  \\
\hline
\multicolumn{7}{c}{SDSS working sample - 22 \mics\ and 250 \mics\ detected} \\
\hline
$0.04<z<0.10$ & 8    & 421  & 51   & 87   & 35   & 602  \\
$0.10<z<0.15$ & 2    & 144  & 35   & 35   & 12   & 228  \\
\hline
\end{tabular}
\end{minipage}
\end{table*}

We classify galaxies in the SDSS sample using the popular emission-line ratio diagram of \othree/\hb\ vs. \ntwo/\ha\
\citep[the BPT diagram;][]{bpt81,vo87}. We follow the method outlined by \citet{kewley06}; curves in the diagram are used to separate
H II galaxies, AGN, and `Composite' systems, which have line ratios intermediate to the first two classes. As a minimum requirement for
a classification using the BPT diagram, we require an \ha\ measurement with a SNR$>3$, satisfied by 2188 objects in the working sample. 
Galaxies with weaker \ha\ are categorized as `Dim'.  If all four lines of the BPT diagram are detected with SNR$>3$ 
in the spectrum of a galaxy ($\approx$35\% of the working sample), the classification is made according to the galaxy's line ratios 
with respect to the curves from \citet{kewley06}. If any of the lines (other than \ha) are detected at a lower significance, a limiting
value for the line is adopted, corresponding to $3 \times \sigma_{line}$, where $\sigma_{line}$ is the uncertainty in the line measurement.
The limit on a particular line is propagated into a limit on the corresponding line ratios that include that line,
and a decision tree is used to identify if the limit on the ratio allows an unambiguous classification of the galaxy. For example, if
\ha, \hb\ and \ntwo\ are well-detected in a galaxy, but \othree\ is weak, the $3 \sigma_{[O III]}$ limit leads to an upper limit on 
\oiii/\hb. If this limit lies below the curve separating H II galaxies from composite systems, the galaxy is unambiguously
classified as an H II galaxy. If it lies above the curve, its classification is listed as `Uncertain'. A similar set of rules are applied to
the other lines or combination of lines. 

We use the term `AGN' exclusively for those systems which show line ratios in their SDSS spectra characteristic of
a hard ionising spectrum. Composite galaxies exhibit lower ionisation levels than AGN only because a majority 
contain an admixture of SF and AGN line emission within the SDSS fibre aperture. We use the broad term `active galaxies' throughout this work
to denote the combined population of AGN and Composites, which include all systems with detectable SMBH accretion. 
Those without clear signatures of high ionisation emission, H II, Dim and Uncertain galaxies, 
are lumped together under the category of `inactive galaxies', though some may host obscured, diluted, 
or faint nuclear activity that is missed by emission line tracers \citep[e.g.,][]{trump15}.

The narrow-line active galaxies in our working sample do not include many luminous systems. 
In the assumption that the extinction-corrected \othree\ luminosity only arises from ionisation by
an accreting SMBH, the bolometric luminosities of the active nuclei are estimated, using a relationship from 
\citet{trump15}, to be between 
$10^{41}$ \ergs\ to $10^{43}$ \ergs, with a median value of $4\times10^{42}$ \ergs.
At these luminosities, the MIR emission from hot dust in a putative torus would not even be detectable in the WISE
22 micron band for more than 90\% of our active galaxies \footnote{This is estimated using the \lothree\--$L_{\rm MIR}$
relationship from \citet{rosario13d} and the mean AGN SED from \citet{mor12}.}. 
We can safely ignore any complications arising from AGN-heated dust in our SED fits.

There is some debate over whether the emission lines in low-ionisation AGN, so-called LINERs, may be primarily excited by
sources unrelated to black hole accretion, such as shocks \citep{heckman80} or hot stars \citep{terlevich85, singh13}.
We acknowledge that the AGN class may encompass a variety of physical processes. In the next paper in this series, which
explores the relationships between star-formation and SMBH accretion, we examine the star-forming properties of LINERs in more
detail.

Table 1 summarises the division of the SDSS working sample into the five classes
considered in this work: Dim galaxies, H II galaxies, Uncertain classifications, composite galaxies and AGN. The sample
is split into two redshift bins to minimise redshift-dependent effects and those due to large-scale clustering (Section 3.1). 

In Figure \ref{bpt_byclass}, we show a pair of BPT diagrams to illustrate the classification method, its results and some properties
of the various populations. Unambiguous H II galaxies, Composites and AGN are shown in the left panel and 
Uncertain galaxies only on the right. Galaxies with a limit on one or both line ratios are indicated by red arrows with directions that show
the nature of the limit. About 61\% of AGN and 32\% of H II galaxies have a limit in one or more BPT line ratio, but can be classified
unambiguously. A requirement of high S/N detections on all four lines in the BPT diagram, frequently adopted in SDSS
population studies, will miss substantial numbers of weak AGN and high metallicity H II galaxies, which we can recover by
a simple method of folding in limits on the line ratios.

About 15\% of galaxies are classified as Uncertain. Of these, 63\% are undetected in both 
the \othree\ and/or \hb\ lines, preventing any constraints along the ordinate of the BPT diagram. 
Their distribution in the \ntwo/\ha\ line ratio is represented by the inverted histograms in the right panel of Figure \ref{bpt_byclass},
where the grey histogram shows the measured line ratio for galaxies with detected [N II] and the red histogram shows the
distribution of upper limits on the line ratio for galaxies with SNR$<3$ in [N II]. Most galaxies with such weak
\othree\ and \hb\ lines have high \ntwo/\ha, implying high metallicities or levels of ionisation.
The rest of the Uncertain class consists mainly of galaxies with limits in [O III] or \hb\ which place them in the Composite
region of the BPT diagram. Taken together, Uncertain galaxies are mostly emission line-weak galaxies with a
mix of star-formation and other ionisation sources contributing to their line spectra, and include many metal-rich galaxies
with strong dust extinction, which preferentially extinguishes the bluer emission lines. When exploring the FIR properties
of galaxies in the remainder of this paper, we treat Uncertain galaxies as a separate heterogeneous population
and examine their properties in the context of the other four classes.

In our classification scheme, Dim galaxies are defined as those with a lack of detectable Balmer line emission. 
However, a small fraction (10\%) of Dim galaxies have \othree\ or \ntwo\ measured with SNR$>3$. This implies
the presence of a faint, high ionization emission line region in these systems. 
While we still include them in the Dim category, we consider their special nature in
the discussion of FIR detection rates in Section 3.1.

\section{Sample Properties}

% ************************* Fig 4 ***********************************
\begin{figure}
\includegraphics[width=\columnwidth]{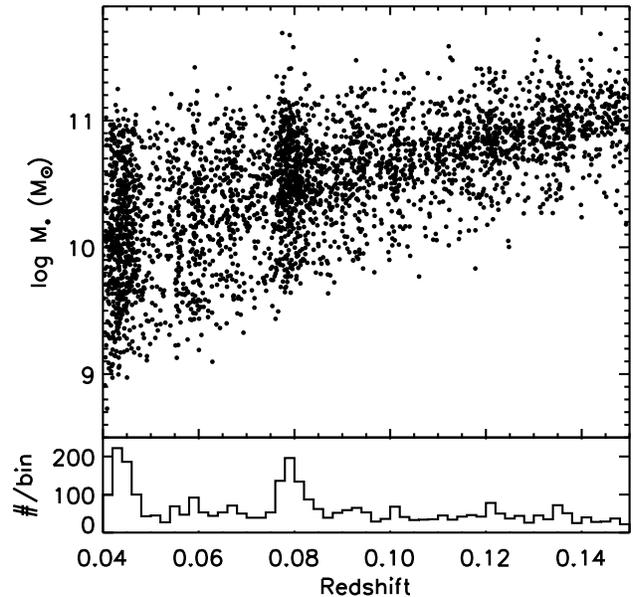}
\caption{Stellar mass plotted against redshift for galaxies from the working sample (upper panel).
The characteristic mass of galaxies increases dramatically towards higher redshifts
due to the magnitude cut of the SDSS Main Galaxy sample. The redshift distribution is
shown in the lower panel. Pronounced spikes are seen from large-scale structure
in the HerS field, particularly at $z<0.1$.}
\label{z_dist}
\end{figure}
% ************************* Fig 4 ***********************************

\subsection{Redshifts and stellar mass distributions}

In the lower panel of Figure \ref{z_dist}, we plot the redshift distribution of the SDSS sample. 
Near-field large-scale structure over the HerS/Stripe82 field can be discerned, in particular the two sharp spikes
at $z\approx 0.045$ and $z\approx 0.08$. Beyond $z=0.1$
the distribution is smoother as the volume occupied by the field gets exponentially larger. The effects of the near-field substructure on
the trends we study in this work through environmental covariances could be important. Therefore, as in Table 1 and when necessary, 
we check our results by splitting the working sample into two fiducial redshift bins: $0.04<z<0.1$ and $0.1<z<0.15$.

The principal selection bias present in the sample relates to stellar mass (\smass). It is evident from the upper panel of Figure \ref{z_dist}
that the range of \smass\ spanned by galaxies in the working sample changes considerably over the redshift range of study due to the
magnitude cut of the SDSS Main Galaxy sample.
Consequently, the characteristic \smass\ of the sample increases by an order of magnitude from $z=0.04$ to $z=0.15$. \smass\
is widely regarded as one of the prime determinants of galaxy properties, and the strong change in its distribution with redshift
has to be taken into account when understanding trends from SDSS-selected galaxy populations.

% ************************* Fig 5 ***********************************
\begin{figure}
\includegraphics[width=\columnwidth]{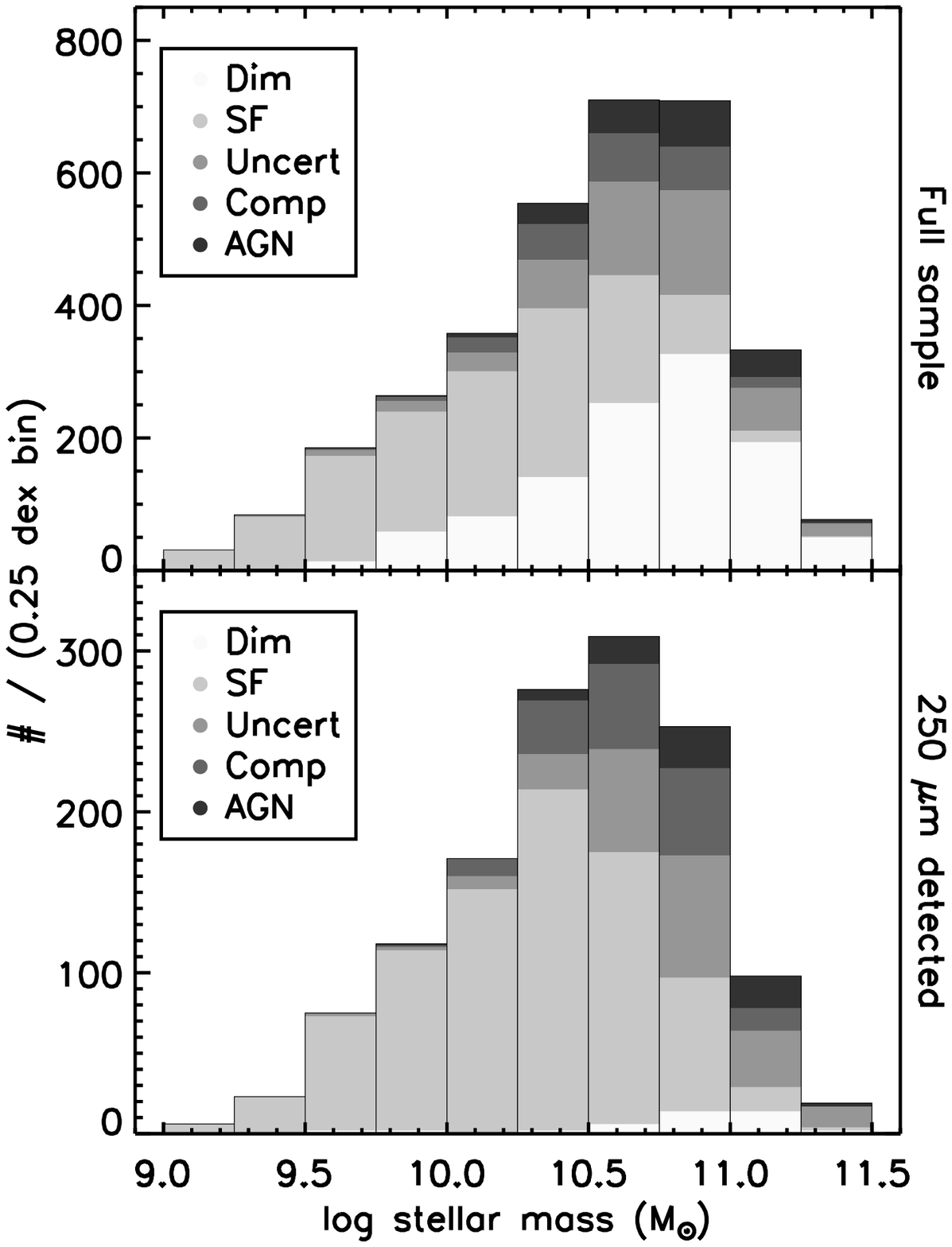}
\caption{{\bf Upper panel:} The break up of SDSS galaxies in the working sample ($0.04<z<0.15$)
into five classes (see Section 2.4) in bins of stellar mass. Different shaded
segments represent the different classes. H II galaxies dominate
at low masses and Dim galaxies dominate at high masses. Composite
galaxies, AGN and galaxies with uncertain classifications have intermediate
masses.
{\bf Lower panel:} Similar to the upper panel, but only for the subset of 
SDSS galaxies detected at 250 \mics\ in HerS. Most Dim galaxies are not detected
in the FIR.
}
\label{main_breakup}
\end{figure}
% ************************* Fig 5 ***********************************

% ************************* Fig 6 ***********************************
\begin{figure*}
\includegraphics[width=\textwidth]{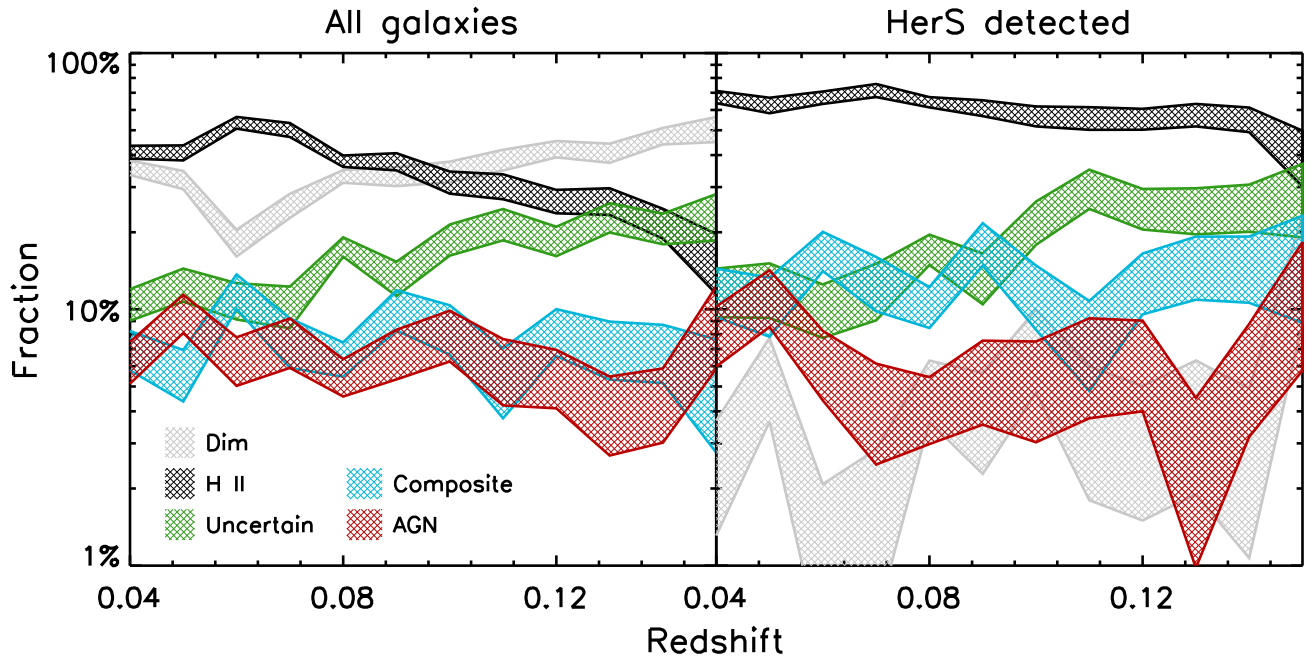}
\caption{The fractions of galaxies of different classes as a function of redshift: Dim galaxies (light grey), H II galaxies (black),
Uncertain galaxies (green), Composites (cyan), AGN (red). The thickness of the shaded bands show the uncertainty 
in the fractions calculated according to binomial statistics in small bins of redshift.
The left panel shows the fractions among all galaxies in the working sample.
The right panel shows only those galaxies detected at 250 \mics\ from HerS.}
\label{z_dist_byclass}
\end{figure*}
% ************************* Fig 6 ***********************************

In the upper panel of Figure \ref{main_breakup}, we graphically present the break-up of the classes of galaxies in our working sample as a function
of \smass. H II galaxies tend to be of low to moderate masses, while Dim galaxies are among
the most massive systems in the sample. AGN peak at \smass$\approx 10^{10.9}$ \msun\ and composite systems are
found at slightly lower masses (\smass$\approx 10^{10.7}$ \msun). Uncertain galaxies have a similar
mass distribution as active galaxies; in terms of gross galaxy properties, they are a closer population
to Composites and AGN than the general subset of H II galaxies.

Further insight is available in the fractions of different classes of galaxies across redshift (Figure \ref{z_dist_byclass}).
In the full working sample (the left panel of the Figure), 
the predominant populations by number are H II galaxies and Dim galaxies, each accounting for 30-40\%
of the total.
Though the majority at $z<0.08$, the fraction of H II galaxies falls with redshift, while Dim galaxies and Uncertain galaxies increase. The
latter class becomes as common as H II galaxies at $z\sim0.13$. This apparent evolution is likely due to a combination 
of the emission line detection limit of the SDSS spectra and the redshift-dependent mass incompleteness of the Main Galaxy sample. 
Towards higher redshifts, an increasing fraction of galaxies have one or more undetected BPT lines, leading to an Uncertain classification,
while more of the low mass H II systems are also excluded. In addition to the smooth decrease with redshift, 
the H II galaxy fraction shows a local peak at $z\approx0.05$, likely due to the environmental effects of large-scale structure
inhomogeneity.

AGN and Composites remain a fairly constant fraction of the population at several percent each across the redshift range. This is because
active galaxies exhibit high \smass\ and do not suffer from strong mass incompleteness at any redshift considered here.

\subsection{250 \mics\ detection rates}

%% ************************* Fig 7 ***********************************
%\begin{figure}
%\includegraphics[width=\columnwidth]{stripe82_class_breakup_250det.eps}
%\caption{The break up of galaxies in the working sample ($0.04<z<0.15$)
%with HerS 250 \mics\ detections. This figure may be contrasted
%with Figure \ref{main_breakup} which shows the full working sample.
%H II galaxies dominate the FIR bright population, while Dim galaxies are essentially
%all undetected in the Herschel maps.
%}
%\label{spire_breakup}
%\end{figure}
%% ************************* Fig 7 ***********************************

%% ************************* Fig 6 ***********************************
%\begin{figure}
%\includegraphics[width=\columnwidth]{stripe82_class_breakup_22+250det.eps}
%\caption{The break up of galaxies in the working sample ($0.04<z<0.15$)
%with detections at both HerS 250 \mics\ and ALLWISE 22 \mics\ bands.} 
%\label{wise+spire_breakup}
%\end{figure}
%% ************************* Fig 6 ***********************************

Applying a HerS detection criterion to the working sample results in a very different mix of galaxy types.
In the lower panel of Figure \ref{main_breakup}, the break-up of the HerS-detected subset is shown as a function of stellar mass
and the fractions of each class across redshift are plotted in the right panel of Figure \ref{z_dist_byclass}.
The most drastic consequence of applying a 250 \mics\ detection is the removal of almost
all of the Dim galaxies. These drop from 34\% of the full working sample to only 3\% of the HerS detected subset.
Even though ionised gas accounts for only a small fraction of the gas mass in galaxies, its presence is closely related
to the radiation field that heats interstellar dust.
Interestingly, of the few HerS-detected Dim galaxies, as many as half are also significantly detected in either \ntwo\ or \othree, 
bolstering a link between the gaseous content of ionised emission line regions and heated dust. It is unclear whether the dust emission
reflects low levels of residual star-formation or is heated by an interstellar radiation field arising from hot evolved stars.
There is evidence from the SDSS that quiescent galaxies on the Red Sequence with LINER emission have
younger stellar populations \citep{graves07}. Low levels of intermittent star-formation could be associated with the small gas reservoirs
visible in some Dim galaxies through these emission line regions.

H II galaxies increase from 37\% of the full sample to 62\% of the HerS subset. 
Composites roughly double, while AGN and Uncertain galaxies remain mostly unchanged in their incidence.
The fraction of H II galaxies also falls with redshift, though, due to the predominance of this class, the drop is not as
pronounced as for the full working sample. 

The very low HerS detection rates of the Dim galaxies and the complementary high occurrence of star-forming galaxies
demonstrates that, at the sensitivity of the HerS survey, most of the FIR emission in inactive (or very weakly active)
galaxies is produced by on-going star-formation. 

\begin{table} 
\centering
\caption{250 \mics\ detection Fractions of AGN and Inactive Galaxies (in percent)}
\begin{tabular}{r@{ }llc}
\hline
AGN & & \multicolumn{2}{c}{Mass weighted control} \\
\hline \hline
Pure AGN: & $36^{+3}_{-4}$ & H II galaxies & $88$ \\
& & Uncertain galaxies & $45$ \\
& & Dim galaxies & $4$ \\
& & All inactive & $35$  \\
\hline
Composite: & $69^{+2}_{-3}$  & H II galaxies & $86$ \\
& & Uncertain galaxies & $41$ \\
& & Dim galaxies & $3$ \\
& & All inactive & $40$  \\
\hline
All active: & $54^{+2}_{-4}$ & H II galaxies & $87$ \\
& & Uncertain galaxies & $42$ \\
& & Dim galaxies & $3$ \\
& & All inactive & $38$  \\\hline
\end{tabular}
\end{table}

The lack of change in the pure AGN detection fraction between the full and FIR detected samples indicates that nuclear
activity occurs both in star-forming and quiescent galaxies. The \smass\ distribution of AGN, 
peaks between those of H II galaxies and Dim galaxies, suggesting that their hosts are drawn from both these disparate populations.
It is well established that the SFR and the fraction of quiescent galaxies in the local Universe is a strong 
function of stellar mass \citep{baldry06,salim07}. Therefore, for a better assessment
of the relative number of star-forming and quiescent galaxies that contribute to the AGN population, we compare the 250
\mics\ detection fractions of AGN to the detection fractions of inactive galaxies after weighting the mass distribution of
the inactive galaxies to match those of the AGN. 

We only consider galaxies in the stellar mass range of $10^{9.5}$ \msun\ to $10^{11.5}$ \msun\ over which AGN are found 
in our working sample. The weighted detection fraction $R_{g}$ of inactive galaxies is calculated as follows:

\begin{equation}
R_{g} \;  = \; \frac{\Sigma n_{g,i} w_{i}}{N_{g} \Sigma w_{i}}
\end{equation}

\noindent where $N_{g}$ is the total number of inactive galaxies in the mass range,
$n_{g,i}$ is the number of inactive galaxies detected at 250 \mics\ in the small bin in stellar mass $i$, and

\begin{equation}
w_{i} \; = \; \frac{n_{g,i}/N_{g}}{n_{a,i}/N_{a}}
\end{equation}

\noindent where $n_{g,i}$ and $n_{a,i}$ are the number of inactive galaxies and AGN from the working sample
in stellar mass bin $i$ and $N_{a}$ is the total number of AGN in the mass range. We adopt a set of independent stellar
mass bins, each 0.25 dex wide, contiguously spanning the stellar mass range. In Table 2, we summarise the detection
fractions for pure AGN, composites and all active galaxies, and compare them to the fractions
calculated for H II galaxies, Uncertain galaxies, Dim galaxies and all three populations of inactive galaxies combined. 

Active galaxies have a significantly higher HerS detection fraction (53\%) than the equally
massive combined population of inactive galaxies (38\%). 
Composites are detected in HerS at twice the rate of pure AGN, 
consistent with their nature as active galaxies with clear contributions to their emission line spectra from
star-forming H II regions. However, even pure AGN are significantly more likely to be detected than inactive quiescent galaxies. 
Therefore, the lack of composite emission line signatures in an active galaxy does not mean that star-formation is absent
in its host. In fact, the detection rate of pure AGN (36\%) is identical to that of the combined inactive population with the same
mass distribution (35\%). This suggests that pure AGN are drawn from the massive galaxy population 
in proportion to the relative incidence of star-forming and quiescent inactive galaxies, while composite systems are strongly
weighted to star-forming hosts.
%We will test
%this view in the following analyses constraining the source of the FIR emission in AGN,
%its relation to star-formation, and by placing the AGN in the context of the general population
%of inactive galaxies.

\section{Multi-wavelength calibrations of the SFR}

% ************************* Fig 2 ***********************************
\begin{figure}
\includegraphics[width=\columnwidth]{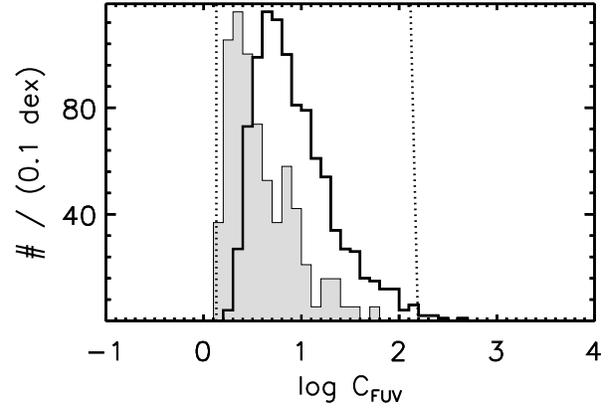}
\caption{Histograms of the dust extinction correction factor in the FUV ($C_{\rm FUV}$; Equation 4)
The open histogram is the distribution for galaxies in our working sample that are detected 
both in the GALEX FUV band and at 250 \mics\ in HerS. The shaded histogram is the distribution
for lower mass, very nearby galaxies from the Herschel Reference Survey. The dotted lines
show the range in $C_{\rm FUV}$ for the sample of normal, star-forming galaxies in \citet{hao11}.
}
\label{fuv_correction}
\end{figure}
% ************************* Fig 2 ***********************************

In the assumption that all the IR luminosity in galaxies arises from the reprocessing of 
optical and UV light from continuously forming stellar populations,
one may use the ratio of the IR and observed FUV luminosities to derive the extinction of
the FUV light. This correction is weakly dependent on the age or the exact form of the 
star-formation history of the stellar populations, or on the extinction law of the dust. 

\citet{hao11} calibrated the IR-based correction to the observed \lfuv, which we have modified slightly
due to the difference in the definition of the integration range for \lir\ between that work
and ours (see Section 2.5 for details). From Equation 15 of \citet{hao11}:

\begin{equation}
C_{\rm FUV} =  1 + 0.50 \times \frac{L_{IR}}{L_{\rm FUV}}
\end{equation}

The scaling term is empirically determined from a local sample of normal star-forming galaxies. 
The intrinsic (i.e., extinction-corrected) FUV luminosity (L$^{c}_{\rm FUV}$) can then be calculated as
the product of $C_{\rm FUV}$ and the observed \lfuv.  

Empirically, as demonstrated by \citet{hao11}, normal star-forming galaxies display a tight relationship 
between L$^{c}_{\rm FUV}$ and the extinction-corrected H$\alpha$ luminosity, with
a scatter of $< 0.1$ dex. In fact, the IR-based correction yields a
more accurate estimate of the extinction-corrected \lfuv\ than the popular approach of using 
the UV slope itself. 

In Figure \ref{fuv_correction}, we plot the distribution of $C_{\rm FUV}$ for the galaxies in the working sample
that are detected both at 250 \mics\ and in the FUV band. They span the full range of $C_{\rm FUV}$
shown by the galaxies in \citet{hao11} (between the dotted lines in the Figure). 
The characteristic value of $C_{\rm FUV}$ from our sample is $\approx 7$, implying that, on average, $>85\%$ of the intrinsic FUV luminosity 
is absorbed. About 23\%\ of FIR-detected sources do not have FUV detections. For these sources, the FUV flux limit at
the threshold GALEX coverage adopted for this analysis (Section 2.4) corresponds to 
$C_{\rm FUV} \gtrsim 150$. For only $\approx 17$\%\ of the FIR-detected sample is the unabsorbed FUV light a 
substantial part ($> 20$\%) of their intrinsic FUV emission. 

The simplest conclusion from the analysis of $C_{\rm FUV}$ is that the escape fraction of FUV light
is quite low among the HerS/SDSS galaxies. Of course, this conclusion is only valid if \lir\ is not strongly
boosted by heating sources that are not related to star-formation, such as AGN or very old evolved stars.
In Section 5.1, we test this assumption and find that most of the FIR can be attributed to star-formation
in our galaxies. This is also consistent with elevated IR luminosities and detection rates among
H II galaxies. 
%In normal star-forming galaxies, as much as 50\%\ of \lir\ may arise in 
%stars older than 100 Myr \citep{hao11}. A scaling down of $C_{\rm FUV}$ by a factor of 2, however, will
%still not change the basic dominance of the obscured and reprocessed luminosity from star-formation
%among the HerS-detected galaxies.

The SFR for galaxies detected jointly in the GALEX/MIS and HerS can be calculated from L$^{c}_{\rm FUV}$ using 
the calibration from \citet{hao11} or \citet{kennicutt12}, valid for a model stellar population with solar abundance
which forms stars continuously for 100 Myr with a \citet{chabrier03} IMF:

\begin{equation}
SFR \textrm{ (UV + TIR)} =  \frac{L^{c}_{\rm FUV}}{2.24\times10^{43}} \; \; \textrm{M}_{\odot}  \textrm{ yr}^{-1}
\end{equation}

A crucial first check is to compare these multi-wavelength SFRs with those derived from the SDSS
for the H II galaxy population. Their SDSS-based SFRs are mostly
estimated from emission-line luminosities emitted by H II regions, widely considered the gold standard for prompt SFR tracers.
In other classes of galaxies, the \dfour\ calibration is used, which can be quite uncertain. Therefore,
we begin by ascertaining the accuracy of our multi-wavelength SFRs in H II galaxies, and apply them
to other classes in following sections.

% ************************* Fig 2 ***********************************
\begin{figure}
\includegraphics[width=\columnwidth]{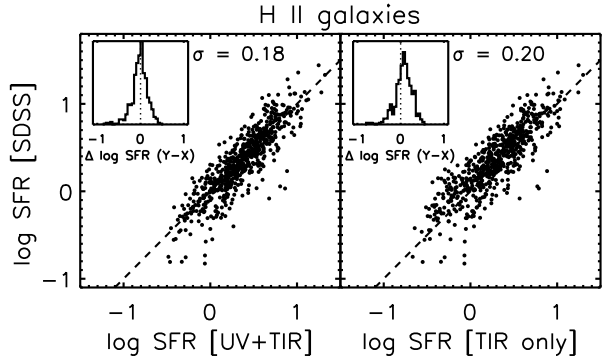}
\caption{A comparison of SFRs from the SDSS (Y-axis) and 
based on multi-wavelength tracers (X-axes) for the H II galaxies
in our working sample. All SFRs are in units of \msun\ yr$^{-1}$.
Dashed lines are 1:1.
In the inset plots, distributions of the ratio of the SFRs (in logarithmic units) 
are shown for the SFRs plotted in each panel. The standard deviation ($\sigma$)
of the distributions are written to the right of the insets.
}
\label{sfr_comps_mw}
\end{figure}
% ************************* Fig 2 ***********************************

In the left panel of Figure \ref{sfr_comps_mw}, we plot the SFRs of H II galaxies from the SDSS against their SFRs estimated from 
multi-wavelength calibrations. The correspondence is excellent, with a uniform scatter about the 1:1 line and a standard deviation of 0.18
dex. This scatter is comparable to the uncertainties on the SDSS SFRs themselves.

Essentially all galaxies in the working sample have high values of $C_{\rm FUV}$, indicating that the bulk of the emission
from young stars goes into the heating of dust. Therefore, we also consider the performance of a pure IR-based SFR calibration. 
Such a calibration has the advantage that it does not rely on the FUV, making it insensitive to contamination from 
non-stellar sources of UV light, in particular faint AGN which may be important in galaxies with accreting SMBHs.  

In the limit that the FUV is completely absorbed by dust, Eqn.~4 \& 5 can be combined to get:

\begin{equation}
SFR \textrm{ (TIR)} =  \frac{L_{IR}}{4.48\times10^{43}}  \; \; \textrm{M}_{\odot}  \textrm{ yr}^{-1}
\end{equation}

The right panel of Figure \ref{sfr_comps_mw} shows the SDSS SFRs of H II galaxies against their SFRs estimated
purely from \lir. Not surprisingly, the correspondence is a bit worse, with a median offset of 0.08 dex (20\%), though with only
a slightly larger standard deviation of 0.2 dex. However, these differences are minor enough for our primary purpose
of evaluating the SDSS-based SFRs of active galaxies. If we adopt a statistical
correction of 0.08 dex  to TIR-based SFRs from Equation 6 for the unabsorbed luminosity from star-formation,
we can proceed to use SFRs derived purely from the TIR 
for HerS-detected galaxies to an accuracy of about a factor of 3. We estimate the error on the TIR-based SFRs as a quadrature
sum of the errors on \lir\ and an additional constant uncertainty of 0.09 dex 
to account for the larger scatter of pure TIR SFRs with respect to SDSS SFRs among H II galaxies.

In Table~\ref{properties_table}, we compile a number of useful properties of all 3319 galaxies in our working sample,
including estimates of \lir\ and pure TIR SFRs.

\begin{table*}
\label{properties_table}
\centering
\caption{Properties of galaxies from the SDSS Main Galaxy sample in the Herschel Stripe 82 survey at $0.04<z<0.15$ (Full version available with the online Journal)}
\begin{tabular}{cccccccc}
\hline
SDSS OBJID & z & Class & \smass\ & SFR [SDSS]$^{\rm a}$ & \lir$^{\rm b}$ & $\alpha$$^{\rm c}$ & SFR [TIR] \\
  &&&(log \msun) & (log \msun\ yr$^{-1}$) & (log \ergs) & &  (log \msun\ yr$^{-1}$) \\
\hline 
588015509277573268 & 0.04494 & Dim        & 10.29 & $-1.89<-1.13<-0.16$ &       -----       &  --   &  --  \\
588015507667419322 & 0.04456 & Dim        &  9.88 & $-2.37<-1.59<-1.06$ &       -----       &  --   &  --  \\
588015507667419317 & 0.04608 & Dim        & 10.17 & $-2.53<-1.71<-1.14$ &       -----       &  --   &  --  \\
587731511538876510 & 0.04068 & H II       &  9.23 & $-0.37<-0.20<+0.06$ & $43.06<43.17<43.22$ & 2.38  & -0.40\\
587731513148047431 & 0.04296 & Dim        & 10.22 & $-2.01<-1.23<-0.70$ &       -----       &  --   &  --  \\
587731512069587131 & 0.05367 & H II       & 10.74 & $+0.35<+0.52<+0.70$ & $43.83<43.92<43.97$ & 2.38  & 0.35 \\
587731512610324707 & 0.04521 & H II       & 10.77 & $+0.51<+0.70<+0.93$ & $44.28<44.29<44.30$ & 2.19  & 0.72 \\
587731513142542498 & 0.04643 & AGN        & 10.61 & $-0.63<-0.04<+0.45$ & $43.83<43.86<43.86$ & 2.06  & 0.29 \\
587731512608292991 & 0.05613 & H II       & 11.13 & $+0.18<+0.52<+0.87$ & $43.87<43.90<43.93$ & 2.75  & 0.33 \\
587731512069455963 & 0.04473 & AGN        & 10.79 & $-1.40<-0.49<+0.01$ & $43.55<43.60<43.64$ & 2.62  & 0.03 \\
587731512069587156 & 0.07735 & H II       & 10.48 & $-0.00<+0.29<+0.61$ & $43.93<43.98<44.00$ & 2.19  & 0.41 \\
587731513144377525 & 0.08156 & Dim        & 10.65 & $-1.78<-0.94<-0.36$ &       -----       &  --   &  --  \\
587731514222510237 & 0.04044 & H II       & 10.05 & $+0.19<+0.40<+0.66$ & $43.86<43.88<43.88$ & 2.12  & 0.31 \\
587731514222248063 & 0.04162 & Dim        &  9.84 & $-2.22<-1.47<-0.97$ &       -----       &  --   &  --  \\
587731514222182633 & 0.04237 & H II       & 10.87 & $-0.20<+0.23<+0.59$ & $43.95<43.97<43.99$ & 2.44  & 0.40 \\
\hline \hline
\multicolumn{8}{l}{\textsuperscript{a}\footnotesize{   16th percentile value $<$ median value $<$ 84th percentile value}}\\
\multicolumn{8}{l}{\textsuperscript{b}\footnotesize{   $1\sigma$ low $<$ best-fit $<$ $1\sigma$ high}}\\
\multicolumn{8}{l}{\textsuperscript{c}\footnotesize{   From \citet{dale02}. See Section 2.5 for details.}}

\end{tabular}
\end{table*}

\subsection{FIR non-detections}

Almost 48\%\ of  galaxies from the working sample that are detected in the GALEX/MIS in the FUV
do not have a 250 \mics\ counterpart in HerS. It is important to understand the nature of these
systems in the overall framework of multi-wavelength star-formation tracers, since this population
will include galaxies with low values of $C_{\rm FUV}$ which are not represented among the FIR-bright
subsample detected in HerS. For example, the distribution of $C_{\rm FUV}$ among the galaxies of the
Herschel Reference Survey \citep[HRS,][]{boselli10} (shaded histogram in Figure \ref{fuv_correction}) is
shifted to a lower median value than either our sample or those from \citet{hao11}. 
The HRS galaxies are primarily low mass systems in the very local Universe (15--25 Mpc away) and include
a number of dwarf galaxies that are not found in the more distant SDSS/HerS sample. Despite this, the
differences in the $C_{\rm FUV}$ distributions in Figure \ref{fuv_correction} certainly indicate that galaxies
with high FUV escape fractions do exist and should be found among the FIR-weak subset.

\begin{figure}
\includegraphics[width=\columnwidth]{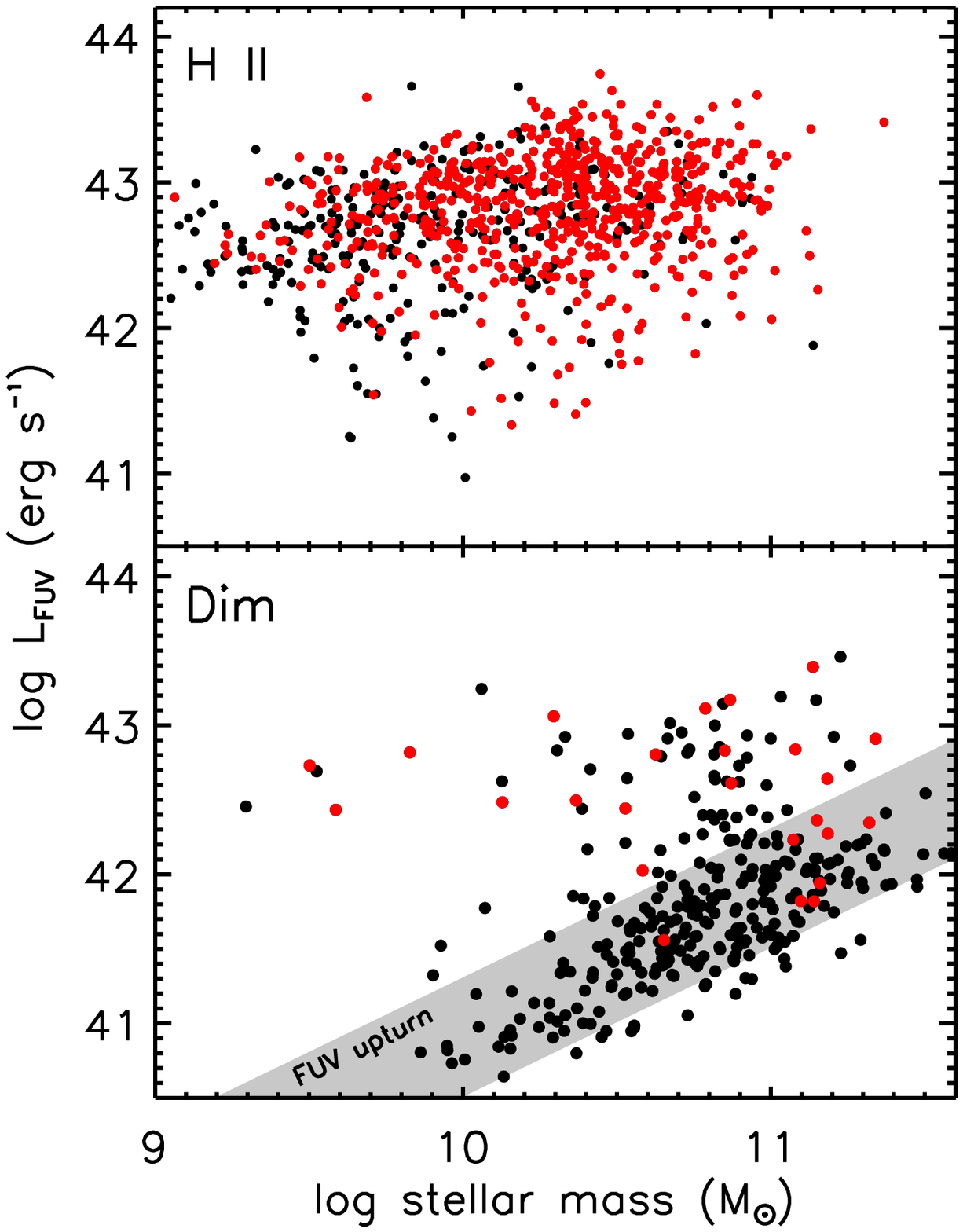}
\caption{The FUV luminosity (\lfuv) plotted against stellar mass (\smass)
of inactive galaxies from our working sample. H II galaxies are shown in the
top panel and Dim galaxies are shown in lower panel. Sources detected
at 250 \mics\ are plotted with red symbols. Clear differences in the properties
of the two populations are evident. Most Dim galaxies have \lfuv\
consistent with the `FUV-upturn' from hot evolved stars, marked
as a grey band in the Figure.
}
\label{fuv_mass}
\end{figure}

A particularly instructive diagram to help us understand the FIR undetected population is shown in Figure \ref{fuv_mass},
in which \lfuv\ is plotted against \smass\ separately for H II galaxies and Dim galaxies from the working sample.
FIR-detected sources are marked with red symbols. 

Among H II galaxies, the fraction with both GALEX and HerS counterparts is high ($\approx 70$\%) -- 
the majority of these galaxies have $C_{\rm FUV} > 3$. Regardless of their FIR properties, H II galaxies
display only a weak correlation between \lfuv\ and \smass\ with considerable scatter.

A surprisingly large fraction of Dim galaxies have FUV detections ($\approx 30$\%), though, in contrast to H II galaxies,
very few of these ($\approx 8$\%) also have FIR counterparts. Since Dim galaxies comprise
the largest subgroup from the working sample, the majority of FUV-detected galaxies that are not
detected in HerS are in fact in the Dim category. Unlike the H II galaxy population, the bulk of Dim galaxies (75\%) show
a fairly narrow linear correlation between \lfuv\ and \smass, though they are typically an order of magnitude
less UV luminous. The small number of Dim galaxies that are bright in the FUV lie in the same region of the diagram and 
show the same trends as H II galaxies. These may be misclassifications due to dust obscuration 
or other effects that affect the visibility of the \ha\ line. 

The low FIR detection rate and weakness of the FUV emission in Dim galaxies suggests 
that the primary FUV-emitting sources in these systems are of a different nature from those in H II galaxies. 
The best explanation is the well-known `FUV upturn' of early-type galaxies, believed to be produced primarily by 
extreme horizontal-branch stars \citep[][and references therein]{oconnell99}. In the lower panel of Figure \ref{fuv_mass},
we show with a shaded band the expected location of galaxies in the diagram due to the FUV upturn. This was calculated 
using the empirical range of $FUV - V$ colours of elliptical galaxies \citep[$2-4$ mag,]{oconnell99} combined with a fixed $V$-band stellar mass-to-light ratio
of a 10 Gyr simple stellar population synthesis model from \citet{han07}. Clearly, the trend seen among most of the 
Dim galaxies is entirely consistent with the FUV upturn phenomenon. The evolved stars that drive the upturn are not related to star-formation
and are not closely associated with ISM dust, which explains the low FIR detection rate of FUV-detected Dim galaxies.

Taking H II galaxies as the representative population of star-forming systems, our analysis of the $C_{\rm FUV}$ distribution
leads us to conclude that, in the range of masses and SFRs shown by our sample, more than 2/3 of galaxies have
low FUV escape fractions. The IR, by and large, is a more complete tracer of the star-forming luminosity of our galaxies.

% ************************* Fig 2 ***********************************
% ************************* Fig 2 ***********************************

\section{SFR comparisons}

% ************************* Fig 2 ***********************************
\begin{figure*}
\includegraphics[width=\textwidth]{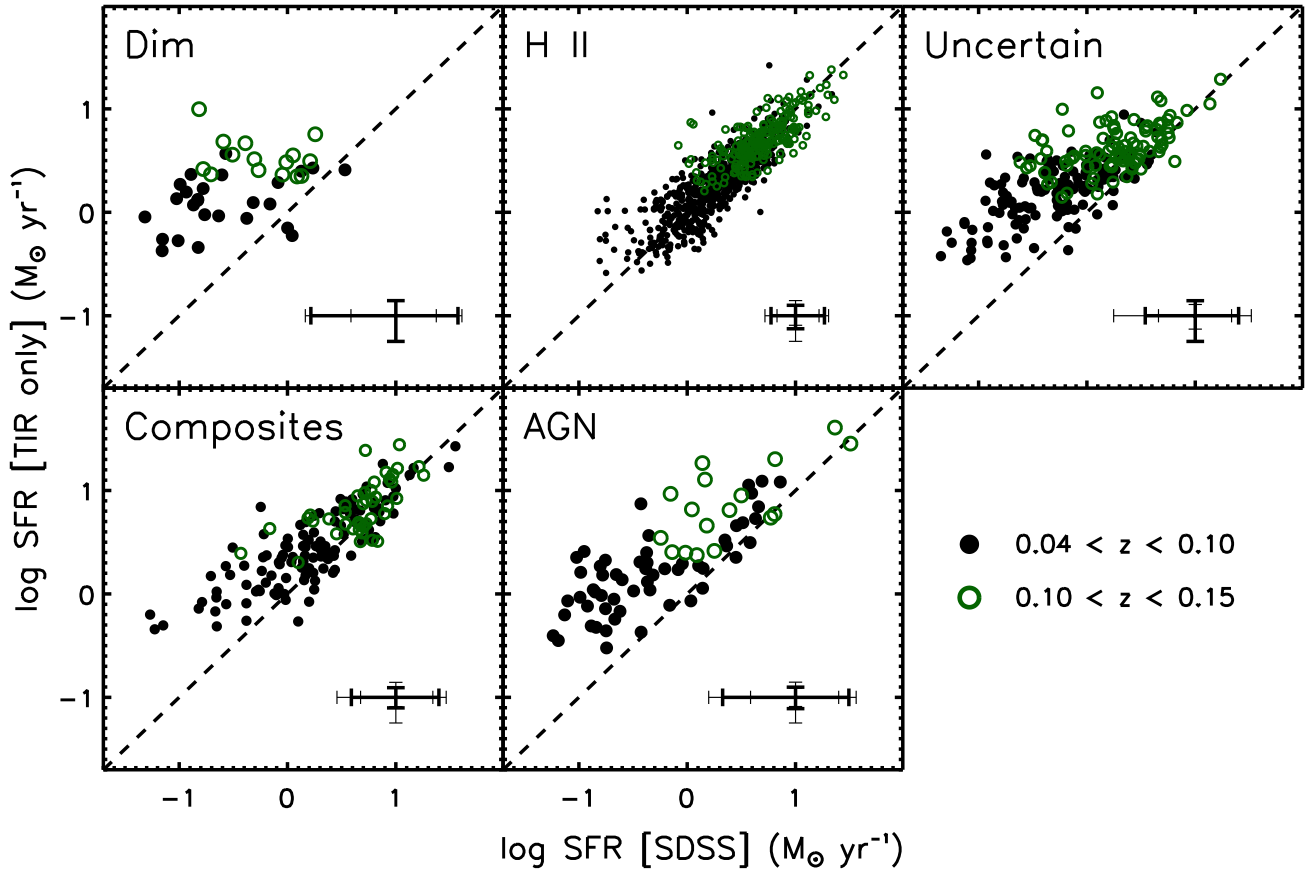}
\caption{Comparison of the SFRs estimated from the IR against the SFRs from the SDSS.
Each panel shows galaxies from different classes (see Section 2.4) 
Two kinds of symbols are used to mark galaxies in the two fiducial redshift bins. 
The bars in the lower right corner of each panel indicate the median errors (thick ticks)
and 16th and 84th percentile errors (thin ticks). Empirical uncertainties on IR SFRs
are smaller and generally more symmetric than the uncertainties on the optical SFRs.
While H II galaxies show a tight 1:1 correspondence between the optical and IR SFRs,
HerS-detected galaxies of other classes show elevated IR SFRs. 
}
\label{lir_comp}
\end{figure*}
% ************************* Fig 2 ***********************************

%% ************************* Fig 2 ***********************************
%\begin{figure*}
%\includegraphics[width=\textwidth]{lir_comparisons_only250.eps}
%\caption{ Same as Figure \ref{lir_comp_bright}, but showing galaxies that are only detected at 250 \mics,
%which are fit using a fixed IR SED template. These galaxies span a range of \lir\ that is lower
%than the galaxies plotted in Figure \ref{lir_comp_bright}, since HerS is more sensitive to fainter IR galaxies
%than ALLWISE.
%}
%\label{lir_comp_faint}
%\end{figure*}
%% ************************* Fig 2 ***********************************

The FIR photometry from HerS allows us to probe down to equivalent IR-derived SFRs $\lessapprox 0.25$ \msun/yr.
This sensitivity and the large sample size available from the HerS area is the main advance in this work.
In the remaining part of the paper, we will use these SFRs to examine optically-derived SFRs from the SDSS/MPA-JHU, 
especially among classes such as active galaxies for which 
well-calibrated emission line indicators of SFR cannot be easily used. 
The assessment of SFR calibrations sets the stage for our second paper, which concentrates on the star-forming
properties of the active galaxies. 
%For our purposes,
%we choose to cast our discussion in terms of comparisons of \lir, rather than the more common approach of converting 
%IR luminosities to SFRs. This is because the IR luminosity, a calorimetric measure of dust heating, can
%arise from sources of energy besides those of star-forming regions. To prevent conceptual confusion and provide a
%simple baseline for possible future studies that build on \lir\ measurements, we instead convert SDSS
%SFRs to an equivalent \lir\ and compare these to our empirical estimates.
%
%For this, we employ well-known conversions for the heating of dust by stellar populations. 
%These conversions depend on the recent star-formation
%history of the galaxy, the distribution of dust and, to some degree, on the grain composition of the dust. 
%Prescriptions for converting SFRs to FIR luminosities typically assume a certain combination of these
%factors. 

In Figure \ref{lir_comp}, we plot the SDSS SFR against TIR-based SFRs (using Equation 6 and a 0.08 dex correction) 
for galaxies from the working sample. For clarity, galaxies are split by classification, and objects in the two fiducial 
redshift bins are shown with different symbols and colors. 

We also show the typical errors on the SFRs for the different classes of galaxies. The errors on 
the optical SFRs are directly from the MPA-JHU catalog. 
For H II galaxies, they reflect the uncertainties on \ha\ fluxes, extinction corrections, 
and corrections to SFRs for the parts of galaxies beyond the SDSS fiber aperture. For all other categories of galaxies, the
errors combine uncertainties on outer SFRs, following the same systematics as H II galaxies, and the 
uncertainties of the \dfour-based calibration of SFRs within the fibre.
Since this calibration has considerable real scatter \citep{brinchmann04}, the SFR errors for the active, Uncertain, 
and Dim galaxies can be considerably larger than those for H II galaxies, even at the same median SFR. 
These errors also tend to be asymmetric, with a larger spread to low SFRs, especially among Dim galaxies.

H II galaxies, which account for most of the HerS-detected galaxies (63\%), are shown
in the middle panel of the top row of Figure \ref{lir_comp} (equivalent to the right panel of Figure \ref{sfr_comps_mw}. 
The close agreement between the two estimates of the SFR is shared by galaxies in both the fiducial redshift bins.
Error-weighted orthogonal bisector regression tests show that there is a small deviation from a slope of unity, but this
is driven primarily by Eddington bias about the HerS flux limit (see Section 5.2). The tightness of the correlation
indicates that most of the TIR luminosity of the H II galaxy population arises in mildly-obscured star-formation
which is also traced by emission lines.

In contrast to H II galaxies, all the other classes show a different behaviour in Figure \ref{lir_comp}.
In general, the IR-derived SFRs for active galaxies (AGN \& composites) and Uncertain galaxies are higher
than their median optical SFRs. The difference is most pronounced for galaxies with low optical SFRs, 
where the offset is $>0.5$ dex. However, even those with higher overall SFRs exhibit
a small net offset of $\approx 0.1$ dex which is not seen among H II galaxies. Composite galaxies show smaller offsets, 
pure AGN and Uncertain galaxies show larger offsets, while the small number of Dim galaxies detected
at 250 \mics\ have the most substantial offsets. In addition, all these classes also exhibit a significantly
wider scatter about a median relationship than H II galaxies, pointing to a weaker correspondence between 
optically-derived and IR-derived SFRs.

The case for Dim galaxies is particularly interesting, since our analysis of their GALEX properties
suggests that a small fraction ($\sim 25$\%) may actually be star-forming galaxies, given their large FUV luminosities.
Among this subset of Dim galaxies, the 250 \mics\ detection fraction (21\%) is much higher than the typical fraction
for the Dim class, which further supports our claim that BPT-based classification in the SDSS misses a non-negligible
population of relatively massive, star-forming galaxies. These misclassified Dim galaxies make up a major
proportion of the points in the top left panel of Figure \ref{lir_comp}.

In the remaining part of this section, we investigate the nature of the offsets seen among some of the HerS-detected SDSS galaxies. 
We use the symbol \dlir $= \log (SFR[\rm TIR] / SFR[\rm SDSS])$ to refer to these offsets in the following discussion.

\subsection{The origin of the SFR offsets}

% ************************* Fig 2 ***********************************
\begin{figure}
\includegraphics[width=\columnwidth]{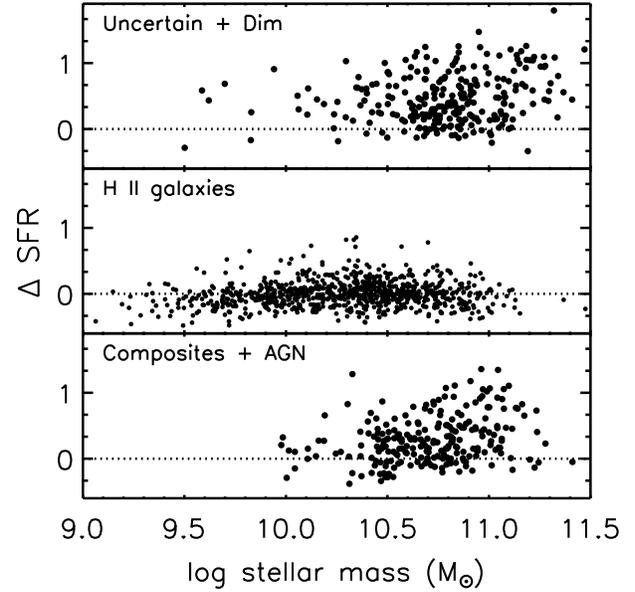}
\caption{ Logarithmic offsets (\dlir) between the IR and optical SFRs 
for galaxies of different classes, plotted against stellar mass (\smass).
The dotted line marks an offset of zero.
}
\label{delta_sfr_mstar}
\end{figure}
% ************************* Fig 2 ***********************************

% ************************* Fig 2 ***********************************
\begin{figure}
\includegraphics[width=\columnwidth]{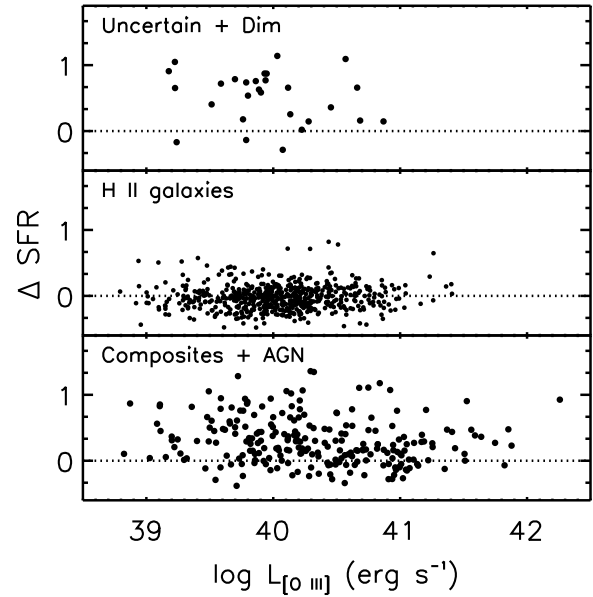}
\caption{  Logarithmic offsets (\dlir) between the IR and optical SFRs
for galaxies of different classes, plotted against the extinction-corrected \othree\ luminosity (\lothree).
The dotted line marks an offset of zero.
}
\label{delta_sfr_lo3}
\end{figure}
% ************************* Fig 2 ***********************************

The offsets between IR-based and optical SFRs could arise because the TIR luminosities of galaxies in 
our working sample have been overestimated, or because their FIR emission contains a contribution from dust that
is not heated by star-forming regions. Alternatively, their SDSS SFRs could be systematically underestimated. 
Considering the large uncertainties of the SDSS SFRs for all classes except H II galaxies (Figure \ref{lir_comp}),
the latter explanation seems most reasonable. Nevertheless, we devise some simple tests to check the other alternatives as well. These
tests look for correlations between \dlir\ and other galaxy properties.

Detailed studies of dust emission in local galaxies have shown that a portion of the long-wavelength FIR luminosity arises in distributed
cold dust which is heated by the diffuse interstellar radiation field \citep{lonsdale87,sauvage92,bendo12}.  
In quiescent galaxies, the interstellar radiation field may arise mostly from
the light of evolved stars which are unrelated to star-formation \citep[e.g.,][]{groves12}. One may postulate that this `cirrus' component
of the FIR luminosity could be dominant in massive galaxies, since most of the stellar mass in such systems is locked in old stellar
populations. If cirrus emission unrelated to current star-formation is the source of excess \lir, 
we expect that \dlir\ correlates with the total stellar mass of galaxies \citep{smith12,boselli12}.

We consider this in the Figure \ref{delta_sfr_mstar}, where we have plotted the \dlir\ as a function of \smass\ for the HerS-detected
galaxies, separating them into three broad categories: H II galaxies, AGN and composites, and the remaining Uncertain and Dim galaxies.
H II galaxies, which do not typically show large \dlir, display no trend over 1.5 dex in \smass. The other classes are more massive and 
show larger offsets. However, there is no significant correlation between \dlir\ and \smass\ for all non-H II galaxies. 
Given that substantial offsets are found over almost an order of magnitude in \smass, 
it is unlikely that the dust-reprocessed light of evolved stars is a leading cause of the excess \lir\ in massive galaxies.

%More insight may be gained by returning to Figure \ref{fuv_mass}. From the lower panel of that Figure, one learns
%that many of the Dim galaxies detected in HerS lie in the part of the \lfuv\ -- \smass\ plane that is occupied
%by star-forming H II galaxies. 

Another explanation is the heating of dust by alternate sources of UV and optical continuum, such as from an AGN or shocks.
These sources also contribute disproportionately to the emission of high ionisation lines such as \oiii, and we may expect, in such a situation,
to find a trend between the extinction-corrected \othree\ luminosity of galaxies and \dlir. As one may gather from Figure \ref{delta_sfr_lo3},
no good correlation is evident between \lothree\ and the offset, either among active galaxies or among 
the few Uncertain and Dim galaxies with detectable \oiii. H II galaxies, as expected, scatter uniformly around a zero offset.

Based on the absence of any significant trends, we conclude that excess cirrus emission or 
heating sources beyond star-formation are not responsible for the bulk of the observed offsets in SFR
that we find in massive non-H II galaxies detected in HerS. 

\begin{figure}
\includegraphics[width=\columnwidth]{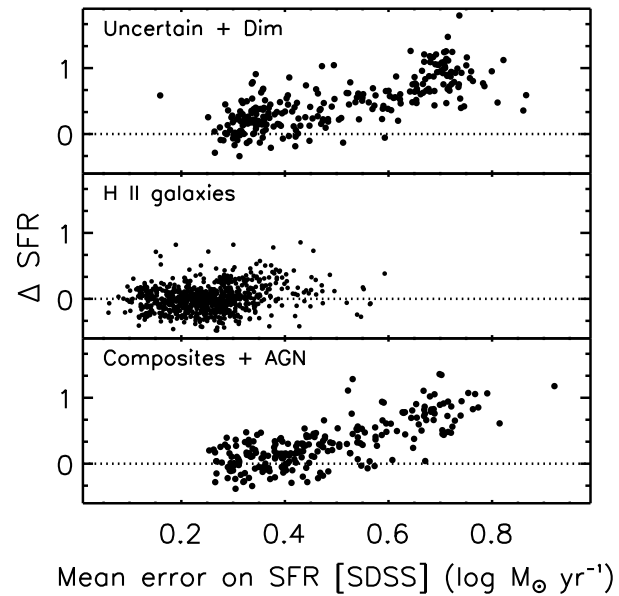}
\caption{ Logarithmic offsets (\dlir) between the IR and optical SFRs
for galaxies of different classes, plotted against the characteristic uncertainty
of the SDSS SFRs. This uncertainty is calculated as half the logarithmic
difference between the 84th and 16th percentile of the SFR likelihood distribution
from the MPA/JHU catalog.
The dotted line marks an offset of zero.
}
\label{delta_sfr_esfr}
\end{figure}

Having found no systematic issues with the IR luminosities, we now consider the SDSS SFRs as a possible source of the offsets seen in 
Figure \ref{lir_comp}. Surprisingly, we find that the strongest dependence of \dlir\ 
is not with any physical property of these galaxies, but with the estimated uncertainty of the
SDSS SFRs. We demonstrate this in Figure \ref{delta_sfr_esfr}. The error on the SDSS SFR for a galaxy is taken to be half the logarithmic
difference (or the logarithm of the ratio) between the 84th and 16th percentile of the SFR likelihood distribution for that galaxy from the MPA-JHU
catalogs. This quantity is equivalent to a 1$\sigma$ uncertainty for the special case of a normal likelihood distribution. 
Both active galaxies and inactive non-H II galaxies (Uncertain and Dim) display a strong correlation between \dlir\ and the error on the
SDSS SFR. This correlation is not evident among H II galaxies, which tend to have smaller SFR uncertainties.  
Almost all galaxies with large \dlir\ also have very uncertain optically-derived SFRs. Conversely, galaxies with accurate optical
SFRs show small \dlir, regardless of classification. 

This behaviour can be understood as the consequence of standard Eddington bias. If a galaxy has a very uncertain SFR measurement, 
the true SFR can differ from the estimated SFR by a considerable amount. If this true SFR is higher than the estimated SFR and 
places the galaxy above the HerS detection limit, it will appear in Figure \ref{delta_sfr_esfr} with a large positive \dlir. On the
other hand, if the true SFR is close to or below the estimated SFR, the galaxy would likely not be detected in the FIR and therefore not
appear in the Figure. The increasing value of the offset with the SFR error is due to the fact that errors on the SFR are strongly anti-correlated with 
the magnitude of the optical SFR.

In the next subsection, we test this explanation by simulating Eddington bias in the SDSS galaxy population taking into account
the redshift-dependent detection limits of the SDSS and HerS surveys.

%If this is the main explanation for the \lir\ offsets, are not due
%to enhanced FIR emission in these systems, but rather because their SDSS SFRs are very uncertain.  
%

% ************************* Fig 2 ***********************************
% ************************* Fig 2 ***********************************

% ************************* Fig 2 ***********************************
\begin{figure*}
\includegraphics[width=\textwidth]{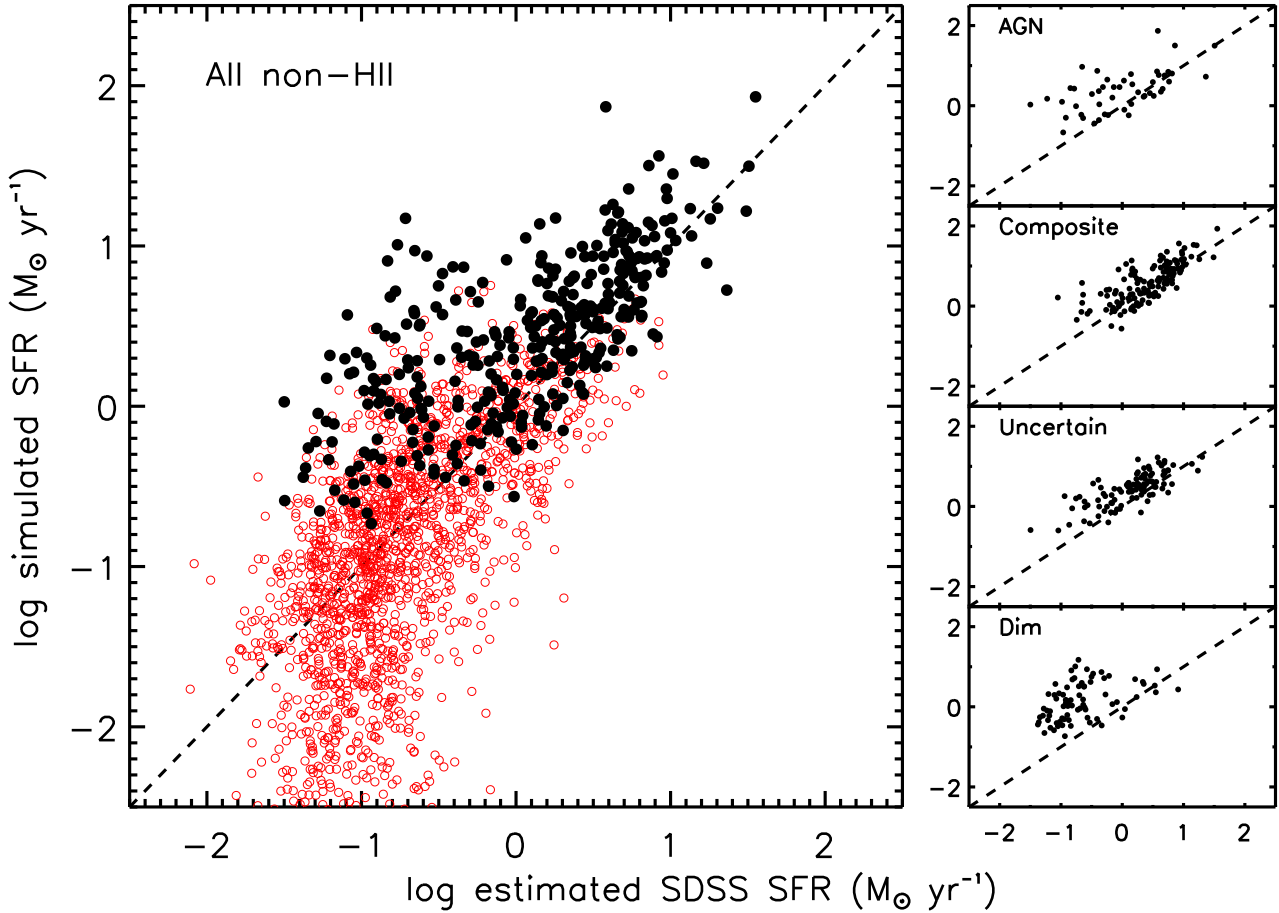}
\caption{Comparison of the simulated SFRs (\sfrs) against
the median SDSS SFRs for Dim galaxies, Composites, AGN and galaxies with uncertain classifications.
All these galaxies (collectively called non-H II galaxies) are shown
in the large panel on the left. Simulated HerS detections (see text for details) 
are shown with solid black points, while simulated non-detections are indicated by open red points. 
In the small panels on the right, the comparison is made for simulated detections only,
with the galaxies separated by their classification. These panels may be compared to
the corresponding panels for each class in Figure \ref{lir_comp}.
}
\label{simulation}
\end{figure*}
% ************************* Fig 2 ***********************************

\subsection{The effects of SFR calibration uncertainties}

The degree and asymmetry of the uncertainties on the optical SFRs, 
combined with the standard redshift-dependent luminosity and volume biases found in the flux-limited SDSS galaxy population,
could potentially cause the large offsets between the observed and expected \lir\ seen in non-H II HerS-detected galaxies. Objects
with the largest asymmetric uncertainties are those with the lowest SFRs. These are only detectable in HerS at lower redshifts. 
When combined with Eddington bias about the FIR flux limit this results in a one-sided scatter about the 1:1 line in 
Figure \ref{lir_comp}. At higher redshifts, Malmquist bias 
restricts us to the subsample with high SFRs, which exhibit more symmetric uncertainties and scatter.
To test this, and to examine the accuracy of the SDSS SFR calibrations, we 
performed a simple simulation of the effects of SFR calibration uncertainties on the correspondence
of optical and IR-based SFRs in HerS-detected galaxies. 

We randomly varied the SFR of each galaxy in our working sample with a probability distribution following the 
calibrated uncertainty of the SFR from the MPA-JHU database, accounting for the asymmetry of the distribution.
This resampled SFR, which we denote as \sfrs, may be thought of as a `true' SFR, which 
differs from the median SDSS estimate of the SFR because of its systematic
uncertainty. \sfrs\ was translated to a TIR luminosity (\lirs) following Eqn.~6. 
For each galaxy, a limiting detectable TIR luminosity \lirl\ was also estimated, as the luminosity of a source 
at the redshift of the galaxy with a randomly chosen SED shape following a probability distribution mirroring the measured $\alpha$ distribution from 
Figure \ref{alpha_dist} and with a 250 \mics\ flux at the HerS detection limit of 30 mJy. This takes into account the 
scatter in SED shapes shown by normal star-forming galaxies. If \lirs$ >$\lirl, the galaxy was flagged as detected in the simulation; 
if not, it was flagged as undetected. One hundred independent iterations of the simulation were performed.

If the likelihood distributions of the MPA-JHU SFR estimates are accurate, 
and if the majority of the TIR luminosity in HerS-detected galaxies is related to on-going
star-formation, then we should expect the galaxies with detectable \lirs\ in the simulation to show the same distribution of \dlir\
as HerS-detected galaxies show in Figure \ref{lir_comp}. Additionally, the 250 \mics\ detection fraction of galaxies in the simulation 
should compare favourably with those empirically measured from HerS.

\begin{figure}
\includegraphics[width=\columnwidth]{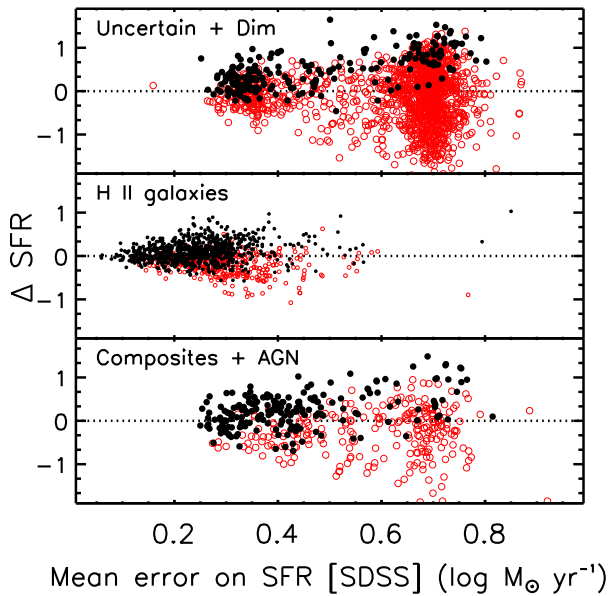}
\caption{The logarithmic difference between the simulated SFR and the median  
SDSS SFR for galaxies of different classes, plotted against the characteristic uncertainty
of the SDSS SFRs. The points plotted here are derived from a single iteration of the simulation described in Section 4.2.
Simulated 250 \mics\ detections are shown as black filled points and non-detections are shown as open red circles. 
This Figure may be compared directly with the empirical version in Figure \ref{delta_sfr_esfr}.
}
\label{delta_sfr_simul}
\end{figure}

In the large left panel of Figure \ref{simulation}, we plot \sfrs\ against the SDSS median SFRs 
of all non-H II galaxies from a single representative iteration of the simulation. 
Since we are interested here primarily in the effects of SFR calibration uncertainties
in systems other than H II galaxies, we only show these populations in this Figure.
With their accurately calibrated optical SFRs, H II galaxies show, as expected, a close correspondence between 
simulated and estimated SFRs (see middle panel of Figure \ref{delta_sfr_simul}, so we exclude them from Figure \ref{simulation} for the sake
of clarity. 

From Figure \ref{simulation}, it is clear that the simulated detections (black points) are offset from the 1:1 line, 
and that this offset depends on the SFR of the galaxy. 
When combined with the non-detections (red points), the overall population shows a more symmetric scatter about the line
of equality, though a small distinct offset is seen throughout. The larger offsets among the detections at lower SDSS SFR is because of
the HerS flux limit, as discussed above. In addition, the spectroscopic flux limit of the SDSS Main Galaxy sample places
a rough lower limit on the SFR of galaxies in the working sample of $\approx 0.03$ \msun/yr. 
This limit results in the apparent vertical edge feature in the simulation at that value along the X-axis.

In the right panels of Figure \ref{simulation}, we split the population by their spectral classifications and only show the simulated
detections. These plots compare favourably with the corresponding panels in Figure \ref{lir_comp}. Another
representation of the behaviour of the different classes is shown Figure \ref{delta_sfr_simul}, in which the simulated \dlir\
from one of the iterations of the simulation is plotted against the optical SFR uncertainty. This Figure mirrors the 
empirical version from Figure \ref{delta_sfr_esfr}, except that it also shows those galaxies with simulated \lir\ 
that would not be detectable in HerS. With increasing SFR error, galaxies display a larger spread in \dlir, producing the
fan-shaped pattern that scatters about the zero offset line. Galaxies with detectable \lirs, shown as black solid points, tend
to have positive \dlir\ and occupy the upper envelope of the pattern, leading to the apparent trends seen among
the actual non-H II galaxy population (Figure \ref{delta_sfr_esfr}). 

Because of their accurate SFRs, calibrated mostly from emission-line luminosities, H II galaxies only show a mild scatter. 
There is a small tail of HerS-detected H II galaxies with low SFRs and modest SFR errors, 
and the action of Eddington bias on this subset can explain the mild deviations from the 1:1 line
seen among the H II galaxy population in Figure \ref{lir_comp}. 

%Much of the qualitative behaviour of each of the classes in that diagram can therefore be attributed quite simply to the differences in the
%SFR calibration uncertainties, without having to invoke additional heating of dust by AGN, shocks or evolved stars. 

%We can also understand the root of the weak correlations of the \lir\ offset with \smass\ and \lothree\ in 
%Figure \ref{delta_sfr_mstar} and \ref{delta_sfr_lo3}. 
%The typical SDSS SFR uncertainty increases with stellar mass, due to the larger fraction of quiescent galaxies at high masses. Among
%active galaxies, it decreases with \lothree, as, due to volume biases, the most luminous composites and AGN are found at 
%higher redshifts and consequently among more strongly star-forming hosts. These other correlations are weak secondary trends driven
%by a strong primary correlation shown in Figure \ref{delta_sfr_esfr}.

% ************************* Fig 2 ***********************************
% ************************* Fig 2 ***********************************

\begin{table} 
\centering
\caption{250 \mics\ detection fractions (in \%): simulated and observed}
\begin{tabular}{rccc}
\hline
Class &  & $0.04<z<0.10$ & $0.10<z<0.15$ \\
\hline \hline
H II & Observed &$66.3_{-1.6}^{+1.5}$ & $78.4_{-2.7}^{+2.3}$\\
       & Simulated & $74.6_{-1.2}^{+1.3}$ & $74.3_{-2.6}^{+2.2}$\\
\hline
Dim & Observed &$3.8_{-0.6}^{+0.9}$ & $3.5_{-0.7}^{+1.1}$\\
        & Simulated & $9.8_{-0.9}^{+1.2}$ & $5.5_{-0.9}^{+0.9}$\\
\hline
Uncertain & Observed &$41.9_{-2.8}^{+2.9}$ & $45.5_{-3.4}^{+3.5}$\\
                & Simulated & $29.6_{-2.3}^{+1.7}$ & $31.3_{-2.9}^{+2.9}$\\
\hline
Composite & Observed &$70.8_{-3.7}^{+3.2}$ & $62.3_{-6.1}^{+5.4}$\\
                  & Simulated & $52.6_{-2.9}^{+2.9}$ & $59.4_{-5.8}^{+4.3}$\\
\hline
AGN & Observed &$36.6_{-3.7}^{+4.0}$ & $32.7_{-5.8}^{+7.0}$\\
        & Simulated & $28.8_{-2.6}^{+2.6}$ & $26.9_{-5.8}^{+3.8}$\\
\hline \hline
\end{tabular}
\end{table}

In Table 3, we list the 250 \mics\ HerS detection fraction of galaxies separated into the five categories and two redshift bins, both from our simulation and 
empirically from HerS. The uncertainties on the simulated fractions are determined from an analysis of all 100 iterations of the simulation. 
The observed detection fractions of H II and Dim galaxies are reproduced by the simulation to within
a few percentage points. Among active and Uncertain galaxies, the simulated detection fractions are generally a 
bit lower than the observed ones, particularly in the lower redshift bin, suggesting that the uncertainties on the optical SFRs for these
classes are underestimated. Alternatively, optical tracers may miss a proportion of heavily obscured
SF in these types of galaxies. We will address this topic in more detail in a forthcoming paper on the molecular gas content of FIR-detected
active galaxies in HerS. The broad consistency between the simulated and observed HerS detection fractions suggests that the SDSS/MPA-JHU
calibrations of SFR that are widely used in there literature are accurate within their substantial uncertainties.

\section{Discussion}

\subsection{Infrared selection of star-forming galaxies}

With the depth and areal coverage of the HerS SPIRE photometry, we are able to detect cold dust emission in the FIR
in $\approx 40$\% of galaxies at $0.04<z<0.15$ from the SDSS Main Galaxy sample in the HerS region.
In contrast, only about 20\%\ are detected in the WISE 22 \mics\ band ($\geq 3\sigma$) \footnote{For the SED fits, we apply a lower threshold for detection
in the WISE W4 band. This is suitable for photometric modeling since the primary selection is a significant detection in the SPIRE 250 \mics\ band. 
In this discussion, we are concerned with an assessment of selection strategy.}. 
In addition, 23\%\ of galaxies are only detected in the FIR with no significant MIR detection, while only 2\%\ are pure MIR sources.
As we demonstrate in Section 2.3 and Figure \ref{mir_fir_color}, there is a population of star-forming galaxies with SEDs dominated
by cool dust which have relatively weak MIR emission. These galaxies are preferentially lost from WISE MIR-selected samples, but are
recovered by Herschel FIR selection. Therefore, MIR-selected studies of star-formation may be systematically biased towards
galaxies with warm dust SEDs. FIR-selected samples are more representative of the full population,
by virtue of selection in a wavelength regime where star-formation heated dust emits most of its radiation.

\subsection{Towards a complete view of star-forming galaxies}

An approach frequently adopted in comparative studies of galaxies in the SDSS and other spectroscopic surveys 
is the use of the BPT diagram to differentiate between galaxies which are star-forming from other populations that
exhibit emission lines, such as active galaxies. However, one or more of the four main BPT emission lines 
may be too faint to detect due to strong dust extinction, particularly among massive and metal-rich
galaxies. Such galaxies have uncertain BPT classifications and are frequently overlooked or neglected in many studies.
These Uncertain galaxies account for $\approx 20$\%\ by number and for $\approx 20$\%\ of the integrated SFR of
massive galaxies ($> 3\times10^{10}$ \msun) in the SDSS. About 40\%\ are detected in HerS, indicating that they 
harbour enough star-formation to place them on or near the Main Sequence of star-forming
galaxies. A full treatment of the star-forming properties of local massive galaxies should take this population into account. The
calorimetric nature of the FIR as a measure of the SFR will allow us to finally do so over the substantial areas of sky covered
by Herschel surveys. In a companion paper, we develop a full context for the frequently-overlooked population of Uncertain galaxies
and show that they show many parallels to the host galaxies of AGN.

\subsection{The reliability of SFR estimates from the SDSS}

H II galaxies show a close correspondence between optical- and IR-derived SFRs (or multiwavelength SFRs), 
which argues that most of the cold dust emission in these systems is heated by SF. All other classes of galaxies, 
if detected in the FIR, exhibit higher TIR luminosities than expected from
extrapolations based on their SDSS SFRs. In these latter categories, SDSS SFRs are evaluated using the optical \dfour\ calibration. 
Our analysis of the uncertainties of their optical SFRs strongly suggests that most of the differences are driven by 
systematics related to these calibrations, in combination with selection effects inherent to the SDSS and the HerS survey (Section 4.2). 
In other words, the apparent enhancements in \lir\ are, for the most part, not due to a dominant contribution from 
obscured SF, AGN-heated dust, or distributed cirrus emission in these galaxies.

The tightness of the relationship between optical SFRs, largely based on prompt emission line-based tracers,
and IR SFRs, which trace dust emission, is quite remarkable. Resolved studies of the dust
emission in galaxies \citep[e.g.,][]{bendo12} have shown that the long-wavelength FIR is frequently more extended
than the short-wavelength emission and does not correlate as well with star-forming regions. This has led
to the suggestion that a substantial component of the cold dust luminosity is associated with a diffuse
interstellar radiation field that could come from evolved stars. Even if the contribution from an old
stellar population was not dominant, the FIR is not a prompt tracer of star-formation since dust can be heated
by stars that are older than a few 100 Myrs \citep{kennicutt09, hao11}. Despite these considerations, the narrowness of the
optical and IR SFR relationship that, among normal star-forming galaxies, the luminosity of cold dust emission
is closely tied to the current level of star-formation. We consider a few explanations that may all play a role.
a) The interstellar radiation field in galaxies may be largely produced by light leaking out from star-forming regions, in which
case the diffuse cirrus emission will also reflect their average SFR. b) The typical star-formation history of galaxies may
be constant for several 100s of Myr. c) The dust emission tracks the mass of dust, which is proportional to the total
molecular gas mass in galaxies, and, through the galaxy integrated Kennicutt-Schmitt law, to the total SFR of galaxies \citep{kennicutt98}.
The last explanation would require a fairly fixed gas-to-dust ratio in galaxies to explain our observations, 
which is at odds with the typical scatter of $>0.3$ dex found even at a fixed galaxy metallicity \citep{remy-ruyer14}.
Regardless of the reason, our study indicates that, among massive galaxies (\smass$>10^{10}$ \msun), the 
TIR luminosity is a reliable measure of the SFR and should be preferred to methods based on age-dependent indicators such as \dfour.

Large area Herschel surveys, such as HerS, can provide accurate FIR-based SFRs for many thousands of galaxies in the nearby
Universe. However, the confusion limit of Herschel maps will prevent direct FIR detections for galaxies fainter than a few 10s of mJy
at 250 \mics. The Main Sequence at \smass$=10^{11}$ \msun\ is detectable only out to $z\sim0.15$.
Across `local' redshifts, Herschel-detected galaxies become increasingly dominated by the rarer star-bursting systems that 
lie well above the Main Sequence. Therefore, Herschel surveys do not provide a full census of SFR, even among local 
massive galaxies. At minimum, the known mass incompleteness of the SDSS and the 
redshift-dependent sampling of the Main Sequence must be taken into account for a consistent treatment of SF galaxies
based on Herschel samples. As a trivial example,
in a monolithic sample consisting of all star-forming galaxies from $z=0.01$ to $z=0.2$ with Herschel detections,
star-bursts will be heavily over-represented compared to the true galaxy population.

%the combination of Herschel-detected galaxies from $z=0.01$ to $z=0.2$ 
%in one monolithic sample results in an inaccurate picture of the relative frequency of star-bursts among local galaxies.

\subsection{The star-forming properties of active galaxies}

As a population, active galaxies are detected at a higher rate in HerS than equally massive inactive galaxies (Section 3.3).
Broadly, nuclear activity is preferentially found in star-forming galaxies, consistent with results for active galaxies selected
by X-ray selection at higher redshifts \citep{silverman09, rosario13b}. While the hosts of nuclear activity are drawn from both 
the star-forming and quiescent populations, as well as quenching systems transitioning between them,
accretion of cold gas on to SMBHs is also linked to the same reservoir of gas that fuels star-formation \citep[e.g.][]{kauffmann07,
rosario13b, vito14}. This is likely the primary link which connects star-formation and nuclear activity, especially among the low
luminosity active galaxies in our sample. Recently, \citet{trump15} show that as many as 30\%\ of active galaxies 
may be missed due to dilution of weak AGN lines by H II regions in the SDSS spectroscopic fibre aperture, implying
that the co-eval connection between active star-formation and AGN may be even more pronounced that we find in this work.

Considering the two classes of active galaxies separately, the `pure' 
AGN are detected in HerS at a rate that is identical to the general population of equally massive 
inactive galaxies ($\approx 35$\%), while the rate for composites is considerably higher ($\approx 70$\%). This is consistent
with the common interpretation of composites as galaxies hosting AGN as well as concurrent star-formation. Table 1 shows
that the HerS detection rate of composites is substantially lower than equally massive H II galaxies, indicating that composites
are not simply H II galaxies with additional nuclear activity but probably include a number of galaxies with lower SFRs than 
the typical H II galaxy population. We examine the relative star-forming properties of H II galaxies, composites and AGN
in the next paper in this series.

\section{Conclusions}

We have undertaken a critical comparison of infrared and optical constraints of the SFR of galaxies using a set of $\approx 3300$
galaxies from the SDSS Main Galaxy sample at $0.04<z<0.15$ covered by the Herschel Stripe82 survey. Our conclusions
are listed below:

\begin{enumerate}

\item The FIR is a reliable measure of the SFR in normal star-forming galaxies (on the Main Sequence).
SFRs derived from the infrared compare well with those from emission line tracers such as \ha. 
The systematic uncertainties on IR-derived SFRs are significantly lower than those from colour or 
spectral index--based methods, making them particularly valuable for active galaxies or dusty galaxies.
Surveys that image the sky to the Herschel/SPIRE confusion limit are capable of detecting galaxies 
on the massive end of the Main Sequence out to z=0.15.

\item The selection of star-forming galaxies based only on the long-wavelength MIR WISE bands is
biased towards systems with warm dust temperatures, and will miss a population of faint  
galaxies dominated by cold dust SEDs. Herschel surveys can recover this population and offers a selection
method that is relatively unbiased to dust temperature variation.

\item Galaxies with one or more undetected BPT emission lines cannot be classified using the standard 
BPT diagram. These galaxies generally exhibit high masses and modest FIR detection rates. 
Their properties overlap considerably with those of active galaxies, and the populations may be related. Such Uncertain
galaxies are frequently ignored or underrepresented in studies that rely purely on emission-line detected samples on the BPT diagram. 
We assert that a more complete picture of star-forming and active galaxies in the local Universe requires that these systems be
fully considered in future studies.

\end{enumerate}

\section{Acknowledgements}
We thank the anonymous referee for a rigorous and constructive review which has substantially improved the quality of this work.
JRT acknowledges support from NASA through Hubble Fellowship grant \#51330 awarded by the Space Telescope Science Institute, which is operated by the Association of Universities for Research in Astronomy, Inc., for NASA under contract NAS 5-26555.
Some of the data presented in this paper were obtained from the Mikulski Archive for Space Telescopes (MAST). Support for MAST for non-HST data is provided by the NASA Office of Space Science via grant NNX09AF08G and by other grants and contracts.
This research has made use of data from the HRS project, a Herschel Key Programme utilising 
Guaranteed Time from the SPIRE instrument team, ESAC scientists and a mission scientist.
The HRS data was accessed through the Herschel Database in Marseille (HeDaM - http://hedam.lam.fr) 
operated by CeSAM and hosted by the Laboratoire d'Astrophysique de Marseille.

\bibliographystyle{mn2e}

\bibliography{mn-jour,hers_stripe82_1}

\appendix

\section{Infrared SED fits: examples and tests}

% ************************* Fig A1 ***********************************
\begin{figure*}
\includegraphics[width=\textwidth]{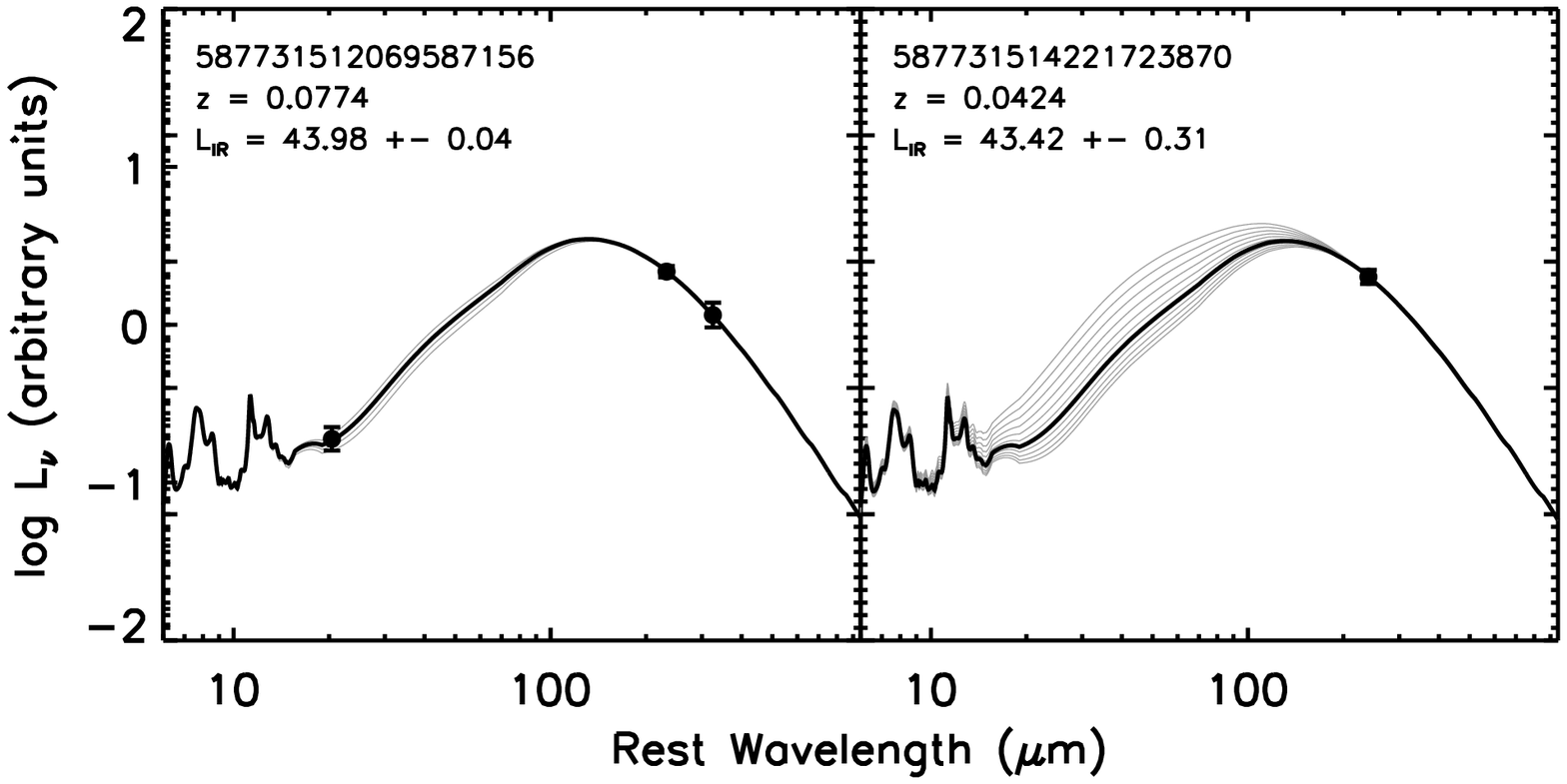}
\caption{Two examples of the SED fits used to estimate \lir\ in this work. In both cases, the best fit SED template is shown
with a solid black line, while the range in templates used to estimate the uncertainty in \lir\ are shown with faint grey lines.
The galaxy in left panel is detected in the Herschel/SPIRE 250 and 350 \mics\ band as well as the WISE 22 \mics\ band. Therefore,
its \lir\ can be constrained very well. In contrast, the galaxy in the right panel is only detected at 250 \mics. The uncertainty
is then estimated from a fiducial range in template shapes (see Section 2.5 for details).
}
\label{example_sedfits}
\end{figure*}
% ************************* Fig A1 ***********************************

In this work, we used WISE and Herschel/SPIRE photometry of galaxies to estimate their total IR luminosity (\lir). Here we briefly examine
the accuracy of \lir\ estimates based only on photometry that does not sample the peak of the IR SEDs of typical galaxies.

% ************************* Fig A2 ***********************************
\begin{figure}
\includegraphics[width=\columnwidth]{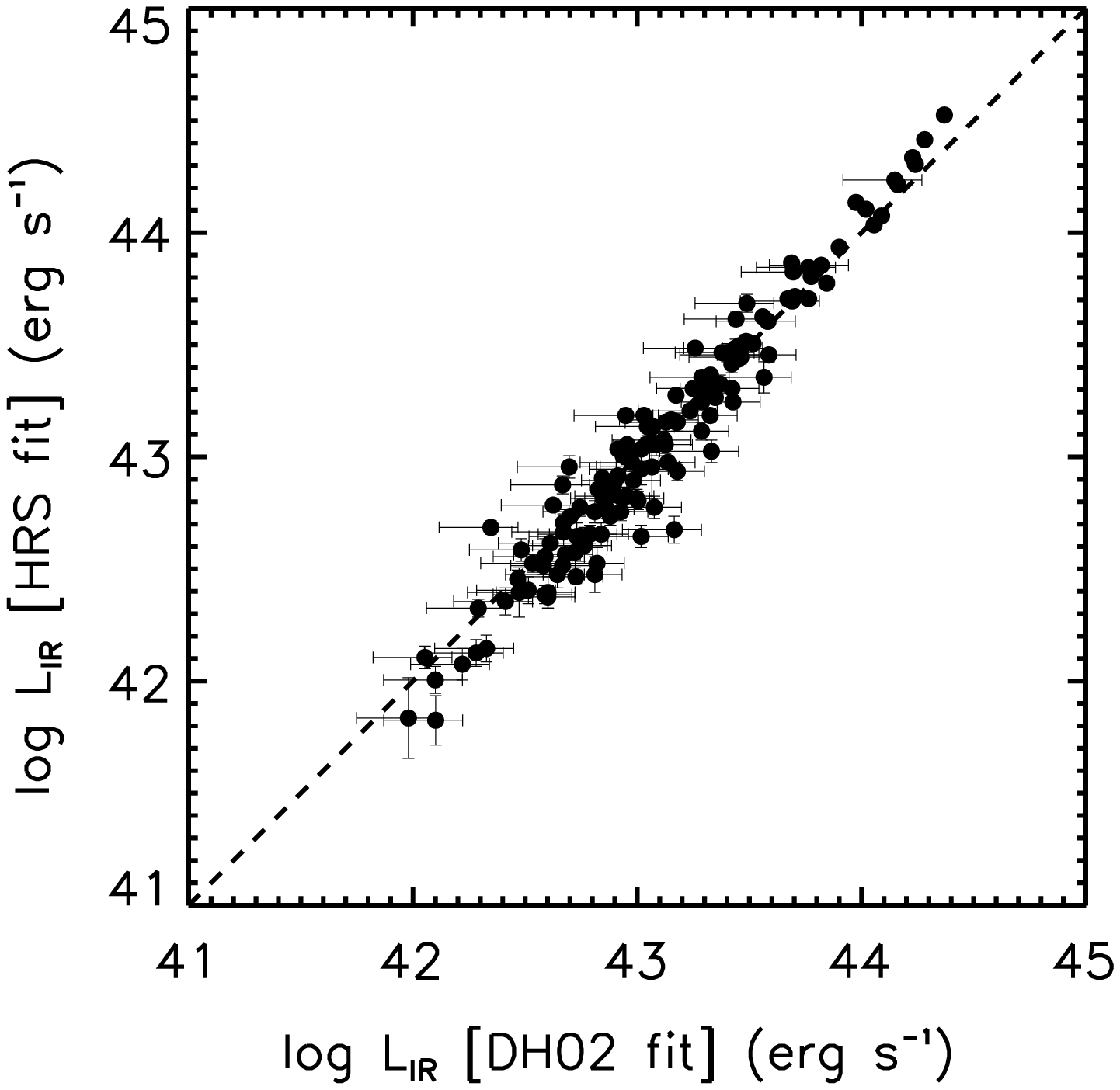}
\caption{ A comparison of \lir\ estimated using our technique (X-axis) and \lir\ derived from a detailed
SED fit to multiple bands (Y-axis), for a set of galaxies from the Herschel Reference Survey (HRS). 
Our technique only fits 24 and 250 \mics\ photometry with the template library of \citet{dale02}, while
the detailed fits, described in \citet{ciesla15}, use all available MIR and FIR photometry and apply
multi-component SED models. 
}
\label{hrs_checks}
\end{figure}
% ************************* Fig A2 ***********************************

In Figure \ref{example_sedfits}, we show two example fits. In both cases, the best-fit template from the \citet{dale02} library is plotted
as a black line, while the grey lines show the range of templates that determine the uncertainties on \lir\ (Section  2.5).
When good photometry is available both in the MIR and FIR, the range in possible galaxy templates is small, as seen in the left panel.
When no MIR photometry is available, the Herschel data itself does not adequately constrain the FIR SED shape. 
In these cases, as shown in the right panel, we adopt a nominal range in galaxy SED shapes to evaluate our uncertainty in \lir.

The accuracy of our \lir\ estimates depends critically on the validity of the \citet{dale02} galaxy template library as a representation
of the real diversity of IR SEDs of galaxies. We test this assumption using a set of 130 local galaxies from the Herschel
Reference Survey \citep[HRS;][]{boselli10} with published IR photometry from Spitzer/MIPS \citep{bendo12} and Herschel/SPIRE
\citep{ciesla12}. We fit the MIPS 24 \mics\ (a close alternative to WISE 22 \mics) and SPIRE 250 \mics\ photometry of these
galaxies using our template-based method. The resulting 3-1000 \mics\ IR luminosities were then compared to estimates
published in \citet{ciesla15}, in which all available MIPS and Herschel photometry of these galaxies, including coverage at 70 and 160 \mics,
were fit with detailed dust models from \citet{draine07}. Figure \ref{hrs_checks} shows this comparison. While more uncertain, our template-based
fits recover the total IR luminosity of these galaxies very well. However, we do find small deviations from the 1:1 line at 
\lir$> 10^{43.5}$ \ergs, at the luminosities that encompass most of our SDSS/HerS sample. Since HRS is a very local survey,
it does not include many luminous and massive galaxies. Nevertheless, the success of our fitting technique in recovering the
\lir\ of most of the HRS galaxies suggests that the \citet{dale02} templates are a suitable way to capture the properties of real
galaxy SEDs when only MIR and FIR photometry beyond the dust peak are available.

\end{document}